\documentclass[
reprint,
amsmath,amssymb,
aps,
prl,
longbibliography,
floatfix
]{revtex4-2}

\usepackage[english]{babel}

\usepackage{graphicx}
\usepackage[T1]{fontenc}
\usepackage[utf8]{inputenc}
\usepackage[table, dvipsnames]{xcolor}		
\usepackage{verbatim}
\usepackage{dcolumn}
\usepackage{hyperref}
\usepackage[capitalise]{cleveref}
\usepackage[mathlines]{lineno}

\usepackage{svg}
\usepackage{siunitx}						
\usepackage{mathtools}						
\usepackage{physics} 						
\usepackage{braket} 						
\usepackage{MnSymbol}

\renewcommand{\i}{\mathrm{i} }
\newcommand{\e}{\mathrm{e}}

\newcommand{\eff}{\mathrm{eff}}
\newcommand{\effcs}{\mathrm{eff,cs}}

\renewcommand{\P}{\mathcal{P}}
\newcommand{\Q}{\mathcal{Q}}

\DeclareSIUnit\fs{fs}						
\DeclareSIUnit\eV{eV}						

\begin{document}

\preprint{APS/123-QED}

\title{Giant counter-rotating oscillations on the attosecond timescale}

\author{Jakob Nicolai Bruhnke
}
\author{Edvin Olofsson
}
\author{Axel Stenquist
}
\author{Jan Marcus Dahlström$^{\text{\dagger}}$
}

\affiliation{Department of Physics, Lund University, 22100 Lund, Sweden. }

\begin{abstract}
\noindent 
We predict an unexplored type of ultrastrong coupling between atoms and intense ultraviolet light that leads to giant population oscillations on the attosecond timescale. These counter-rotating oscillations can be of similar amplitude as the elementary femtosecond Rabi oscillations between the two strongly coupled states. The effect, which is beyond the two-level atom, is non-reciprocal: It only affects the excited state, while the ground state is unaffected. We propose that two-photon Rabi oscillations (1s$^2$--1s3d) in helium is suitable for the generation of this type of ultrastrong coupling with realistic pulses. We use a combination of Floquet theory and effective Hamiltonian theory to test our predictions against \textit{ab initio} simulations.
\end{abstract}
  
\maketitle

\begin{table}[b!]
\begin{flushleft}
  $^\text{\dagger}$ marcus.dahlstrom@fysik.lu.se
\end{flushleft}
\end{table}

\twocolumngrid

The rotating wave approximation (RWA) is an important simplifying assumption in light matter interaction because it helps to circumvent quantum fluctuations by enforcing energy conservation between atoms and photons \cite{cohen-tannoudjiAtomPhotonInteractionsBasic1998}. Understanding the breakdown of the RWA is at the foundation of diverse research fields, such as ultrafast strong-field physics \cite{krauszAttosecondPhysics2009} and cavity quantum electrodynamics (cQED) \cite{friskkockumUltrastrongCouplingLight2019}. Historically, counter-rotating (CR) effects in two-level systems were explored in pioneering works by Bloch and Siegert \cite{blochMagneticResonanceNonrotating1940}, and Autler and Townes \cite{autlerStarkEffectRapidly1955}. In further work, Shirley introduced Floquet theory as a limit of quantum optics, with infinite photon numbers and cavity volume \cite{shirleySolutionSchrodingerEquation1965}, to derive phenomena beyond the RWA in the form of an induced energy shift (Bloch-Siegert shift) \cite{blochMagneticResonanceNonrotating1940}, as well as sub-cycle modulations of the atomic populations (CR oscillations). Due to limits of laser technology at the time, Shirley dismissed these CR oscillations as unobservable, and calculated averaged transition probabilities that did not depend on the instantaneous electric field. In the same spirit, weak and rapid fluctuations of populations that occur in exact propagation of the time-dependent Schrödinger equation (TDSE) for hydrogen atoms in strong fields, have -- if at all -- been mentioned in passing \cite{lagattutaAbovethresholdIonizationAtomic1993, dorrTimeEvolutionTwophoton1997, kaiserPhotoionizationResonantlyDriven2013}. 

In cQED, strong coupling effects are studied using the quantum Rabi model \cite{xieQuantumRabiModel2017}. Here, ``CR effects'' implies a breakdown of the Jaynes-Cummings model \cite{jaynesComparisonQuantumSemiclassical1963}, with eigenstates being composed of energy non-conserving uncoupled atom-field states in the ``ultrastrong coupling'' regime \cite{forn-diazUltrastrongCouplingRegimes2019, friskkockumUltrastrongCouplingLight2019, rossattoSpectralClassificationCoupling2017}. Another breakdown of the RWA occurs in non-resonant interactions between atoms and intense low-frequency laser fields. Over the last decades, semi-classical treatment of such interactions has revealed novel phenomena, such as high-order harmonic generation \cite{lewensteinTheoryHighharmonicGeneration1994} and formation of attosecond pulses \cite{paulObservationTrainAttosecond2001}, which have found numerous applications in atoms \cite{dahlstromIntroductionAttosecondDelays2012}, molecules \cite{calegariChargeMigrationInduced2016} and solids \cite{vampaTheoreticalAnalysisHighHarmonic2014, yueIntroductionTheoryHighharmonic2022, pazourekAttosecondChronoscopyPhotoemission2015}. Attosecond technology has also enabled the study of sub-cycle effects of bound states, and CR oscillations in the time domain, as recently shown by polarizing Rydberg states with infrared fields \cite{chiniSubcycleAcStark2012, reduzziPolarizationControlAbsorption2015, anandAttosecondCounterrotatingwaveEffect2017}. %
Recently, the research communities of quantum optics and strong field physics have begun to intersect with researchers considering the generation of quantum states driven by strong fields \cite{stammerQuantumElectrodynamicsIntense2023, lewensteinGenerationOpticalSchrodinger2021, tsatrafyllisHighorderHarmonicsMeasured2017, gorlachQuantumopticalNatureHigh2020}.

\begin{figure*}[t]
    \centering
    \includegraphics[width=\linewidth]{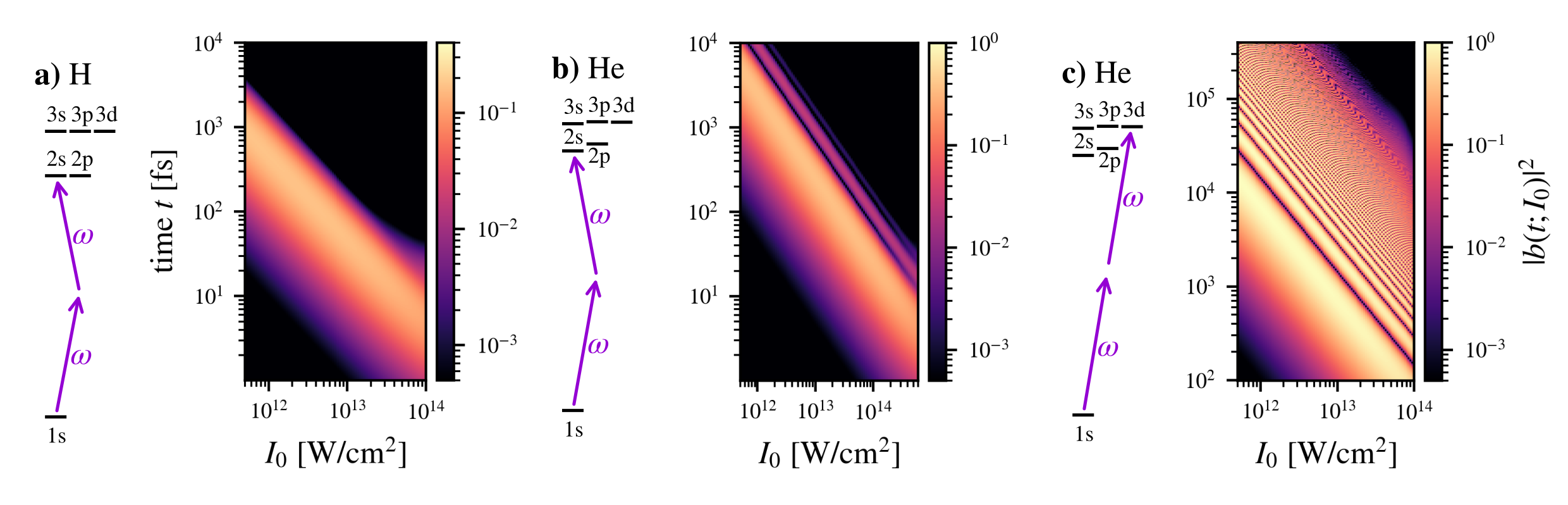}
    \caption{Excited state population $|b(t)|^2$ for the resonant transition (i.e. adjusted for AC Stark shift) from the ground state to (a) the 2s state in hydrogen, (b) the 1s2s state in helium, (c) the 1s3d state in helium, parametrically for different intensities. In analogy to classical mechanics, we can define a damping ratio $\zeta$ in order to classify the overdamped ($\zeta > 1$) and underdamped ($\zeta < 1$) regime. In (a) $\zeta \sim 1.5$, in (b) $\zeta \sim 0.4$, and in (c) $\zeta \sim \num{6e-3}$.}
    \label{fig:poprabi}
\end{figure*}

Strong coupling implies first and foremost Rabi oscillations. In this work, we primarily focus on two-photon Rabi oscillations (2-PRO). 2-PRO is a counterintuitive quantum process that allows for strong coupling between two atomic states of equal parity via weakly coupled states of opposite parity. The Rydberg physics community has diligently applied 2-PRO because it enables coherent transitions to highly excited metastable states in atoms using available lasers in the optical domain \cite{kumlinQuantumOpticsRydberg2023}. As it is a non-linear process, which requires high intensities sustained for long times, the process has been largely unexplored in the UV and XUV regime. One notable exception is the work of Dörr et al., where it was shown that population inversion through 2-PRO in hydrogen atoms between the 1s and 2s states is impossible due to unavoidable ionization losses induced by the laser field \cite{dorrTimeEvolutionTwophoton1997}. In cQED terminology, large losses compared to the Rabi period lies in the ``weak coupling'' regime, $\Omega < \Gamma$, with ionisation rate $\Gamma$ and Rabi frequency $\Omega$. In Fig.~\ref{fig:poprabi}~(a), we reproduce this important and unexpected result by showing that the 2-PRO of the excited 2s state in hydrogen are overdamped in the entire resonant parameter space (spanned by pulse intensity $I_0$ and duration $t$). Physically, this effect arises due to the similar scaling of both $\Omega$ and $\Gamma$, $(\Omega,\Gamma)\propto E_0^2$, as can be understood from effective Hamiltonian theory \cite{beersExactSolutionRealistic1975, holtTimeDependencesTwo1983}. While 2-PRO in the UV and XUV regime have not yet been realized, intense XUV fields produced by seeded free-electron lasers (FEL) \cite{emmaetal.FirstLasingOperation2010, allariaetal.HighlyCoherentStable2012, mirianetal.GenerationMeasurementIntense2021} have recently been used to demonstrate one-photon Rabi oscillations (1-PRO) \cite{nandietal.ObservationRabiDynamics2022} and strong-field quantum control \cite{richteretal.StrongfieldQuantumControl2024} between the atomic ground state (GS) and excited states (ES) in helium atoms. 
In these experiments, $\Gamma < \Omega$, since $\Omega\propto E_0$ and $\Gamma\propto E_0^2$  \cite{olofssonPhotoelectronSignatureDressedatom2023}. Therefore, we propose using the ``strong coupling'' nomenclature from cQED for this regime. 

In this letter, we show that two-photon Rabi cycling helium atoms, driven by intense XUV light, is a promising system to explore light-matter coupling in novel domains. The {\it overdamped} nature of the 1s-2s hydrogen UV transition [Fig.~\ref{fig:poprabi}~(a)], does not generalize to the experimentally relevant helium atom because the 1s$^2$-1s2s XUV transition is {\it underdamped}, albeit with only few weak Rabi oscillations surviving, as can be observed in Fig.~\ref{fig:poprabi}~(b). Most importantly, the two-photon coupling between 1s$^2$ and 1s3d state is underdamped, with strong Rabi cycles accessible in the parameter space, as shown in Fig.~\ref{fig:poprabi}~(c). 
Therefore, the 1s$^2$-1s3d transition allows for ``strong coupling'' conditions over a broad range of intensities and pulse lengths, which enables us to increase the field intensity to explore physics beyond the RWA \footnote{While we will not present results on the following, we should briefly mention that we have found that 2-PRO can indeed be driven in hydrogen between the 1s and 3d. The degeneracy of the 3s and 3d is lifted by the field, allowing one to selectively target the 3d state \cite{bruhnkeTwophotonRabiOscillations2024}}. In analogy with cQED, we will refer to this regime as ``perturbative ultrastrong coupling''. In traditional cQED, which is concerned with ideal two-level systems, the ultrastrong coupling regime is reached when the coupling is comparable to the optical frequency: $\Omega\approx \omega$. In our work, which concerns real multi-level atoms, we find that giant CR oscillations take place already for couplings much smaller than the optical frequency: $\Omega \ll \omega$. Instead, the question that must be posed is how strongly the ES couples to nearby Rydberg states. We find that the giant CR effects can break the reciprocality of ultrastrong coupled states, by which we mean that CR oscillations may appear solely on one of the strongly coupled states. Providing a quantitative explanation of non-reciprocal CR oscillations is the main point of this letter. Atomic units are used unless otherwise stated: $e=\hbar=m=1/4\pi\epsilon_0=1$.

\textit{Theory}---In the following, we provide the outline for the derivation of the CR population dynamics for the two-photon Rabi cycling case. The theoretical modeling of 2-PRO in atoms is enabled through an essential states formalism. The essential states span a sub-Hilbert space $\P$, and they are coupled effectively via the orthogonal Hilbert space of non-essential states, called $\Q = \P^\perp$. Let us refer to the essential states as $\ket{a,0}$ (GS) and $\ket{b,-2}$ (ES), where $a$ and $b$ refer to the ground and excited atomic states, respectively. The second index labels ``photon numbers'' of the mode. Unlike quantum optics, the absolute photon number is simply an inconsequential energy offset in Floquet theory. Therefore, the photon numbers may take on negative values. Parametrizing the effective coupling of the essential states to and via the $\Q$-space leads to an effective two-level Hamiltonian $H_\eff$. If the $\Q$-space contains CR states, $H_\eff$ incorporates CR effects, such as the Bloch-Siegert shift and CR contributions to effective couplings. Since ionization in strong high-frequency fields can not be neglected, the effective Hamiltonian must be non-Hermitian, and photoionization can occur in a range of different non-linear processes \cite{beersExactSolutionRealistic1975, holtTimeDependencesTwo1983, olofssonPhotoelectronSignatureDressedatom2023}. However, the two-level $H_\eff$ generates only a {\it cycle-averaged} time-evolution with populations $|a(t)|^2$ (GS) and $|b(t)|^2$ (ES). It can thus not be used to model rapid CR oscillations. In Floquet theory, CR oscillations of an atomic state $\ket{b}$ stem from interference of the essential uncoupled atom-photon state $\ket{b,-2}\in \P $ with Floquet states comprised of the same atomic state, but a different number of photons, $\ket{b,k}\in \Q$, where $k\ne -2$. 

\textit{Method}---The non-Hermitian effective Hamiltonians employed in this work are obtained through non-perturbative time-independent calculations, enabled through Floquet theory \cite{chuFloquetTheoremGeneralized2004}. We construct a complex-scaled Floquet Hamiltonian $H_F^\theta$, whose continuous spectrum has been rotated into the complex plane \cite{chuIntenseFieldMultiphoton1977, maquetStarkIonizationDc1983}. From $H_F^\theta$, we parametrize the influence of the $\Q$-space states in an all-order approach \cite{suzukiDegeneratePerturbationTheory1983}. Specifics on effective Hamiltonian theory can be found in the End Matter. $H_F^\theta$ is constructed in the length gauge from complex-scaled energies and transition dipole moments, both obtained using the configuration-interaction singles (CIS) method \cite{foresmanSystematicMolecularOrbital1992, dreuwSingleReferenceInitioMethods2005}, with exterior complex scaling (ECS) \cite{simonDefinitionMolecularResonance1979}. To validate the results, we perform \textit{ab initio} simulations using the time-dependent configuration-interaction singles (TDCIS) method in the length gauge  \cite{rohringerConfigurationinteractionbasedTimedependentOrbital2006}. The simulations are done with a CIS basis that is constructed using a B-spline basis with a complex-absorbing potential (CAP) \cite{greenmanImplementationTimedependentConfigurationinteraction2010, bertolinoThomasReicheKuhnCorrectionTruncated2022}. The use of length gauge is required: As shown by Kobe (1978), the complex amplitudes of the states are probability amplitudes only if the length gauge is employed \cite{kobeGaugeInvariantFormulation1978}. This means that their squares correspond to the probabilities to find electrons in the respective field-free states. 

\textit{Analytical formulation}---In Floquet theory, the transition amplitude between the atomic states $\ket{a}$ and $\ket{b}$ from $t=0$,  
\begin{equation}
    U_{ba}(t) = \sum_k \braket{b,k | \e^{-\i H_F^\theta t} | a,0} \e^{\i k \omega t}, \label{eq:timeevolutionoperator}
\end{equation}
is given as the sum over the transition amplitudes between the atom-photon states $\ket{a,0}$ and $\ket{b,k}$, determined by $H_F^\theta$, and weighted with the Fourier components $\exp(\i k \omega t)$ \cite{shirleySolutionSchrodingerEquation1965, chuFloquetTheoremGeneralized2004}. In our essential-states approach, only one photon block ($k=-2$) is included. Expanding in the dressed state basis of $H_\mathrm{eff} \ket{\pm} = \lambda_\pm \ket{\pm}$, we obtain 
\begin{equation}
    U_{ba}(t) \approx \sum_{j \in \pm} \braket{b,-2| j } \e^{-\i \lambda_j t} \Braket{\tilde j|a,0} \e^{-2\i \omega t}.
\end{equation}
The transition probability is the optical cycle-averaged ES population, $|U_{ba}(t)|^2 = |b(t)|^2$. Since $H_\eff$ is non-Hermitian, we require the bi-orthonormal $\bra{\tilde \pm}$, which fulfill the left eigenvalue equation for $H_\mathrm{eff}$, so that $1 = \ketbra{+}{\tilde +} + \ketbra{-}{\tilde -}$ \cite{moiseyevNonHermitianQuantumMechanics2011}. For a brief introduction of Floquet theory, its application to the semiclassical two-level system, and expressions for the damped, effective two-level dynamics, we refer to the Supplemental Material \cite{SeeSupplementalMaterial} and references \cite{guerinRelationCavitydressedStates1997, shoreTheoryCoherentAtomic1990} therein. 

According to Eq.~\eqref{eq:timeevolutionoperator}, CR oscillations are induced when several $\ket{b,k}$ -- in our case $\ket{b,0}$, $\ket{b,-2}$, and $\ket{b,-4}$ -- have non-negligible overlap with the full eigenstates of $H^\theta_F$, i.e., $H_F^\theta \ket{\phi_\pm} = \lambda_\pm \ket{\phi_\pm}$. However, diagonalization of $H_\eff$ only yields the eigenstates $\ket{\pm}\in \P$, while $\ket{b,k}\in \Q$, $k\neq -2$, thus requiring the knowledge of $\ket{\phi_\pm}$. The operator that produces the $\Q$-space components of a given $\P$-space state is known as the {\it reduced wave operator} $\chi$ \cite{killingbeckBlochWaveOperator2003, lindgrenAtomicManyBodyTheory1982}. With $\chi$, we can obtain $\ket{\phi_\pm} \in \P\oplus \Q $ from $\ket{\pm} \in \P$ simply via
\begin{equation}
    \ket{\phi_\pm} = (P + \chi) \ket{\pm}, \label{eq:fullwavefunctionfromP}
\end{equation}
where $P \coloneqq \ketbra{a,0}{a,0} + \ketbra{b,-2}{b,-2}$ projects on the essential states. A sufficiently accurate approximation to $\chi$ naturally appears in our calculation of $H_\mathrm{eff}$, as shown in the End Matter.

Through Eq.~\eqref{eq:fullwavefunctionfromP}, the transition amplitude in Eq.~\eqref{eq:timeevolutionoperator} can be expressed in a form that only involves quantities accessible through effective Hamiltonian theory, 
\begin{equation}
    U_{ba}(t) = \sum_{k \in \{0,-2,-4\}} \sum_{j \in \pm} \braket{b,k| P+\chi |j } \e^{-\i \lambda_j t} \Braket{\tilde j|a,0} \e^{\i k\omega t}, \label{eq:cr_amp}
\end{equation}
but retains sub-cycle dynamics. Interference between the CR states, and coupling between the GS and the $\Q$-space can be neglected ($\chi \ket{a,0} \approx 0$) to obtain a compact form for our transition probability: 
\begin{multline}
    |U_{ba}(t)|^2 = |b(t)|^2 \bigl[1 + |\Lambda^{(0)}|^2 + |\Lambda^{(-4)}|^2 \\ + 2\Re(\Lambda)\cos(2\omega t)  \bigr]. \label{eq:crpopb2} 
\end{multline}
Here, $\Lambda^{(k)} \coloneqq \braket{b,k|\chi|b,-2}$, and thus $|\Lambda^{(k)}|^2|b(t)|^2$ represents the optical cycle-average population of the CR state $\ket{b,k}$, %
$k=0,-4$. Further, $\Lambda \coloneqq \Lambda^{(0)} + \Lambda^{(-4)}$ is the characteristic quantity determining the magnitude of the CR oscillations. The maximum span of the CR oscillations is given by 
\begin{equation}
    M_{\max} = 4|\Re(\Lambda)| \max(|b(t)|^2). \label{eq:Mmax}
\end{equation}

\begin{figure}[h]
    \centering
    \includegraphics[width=\linewidth]{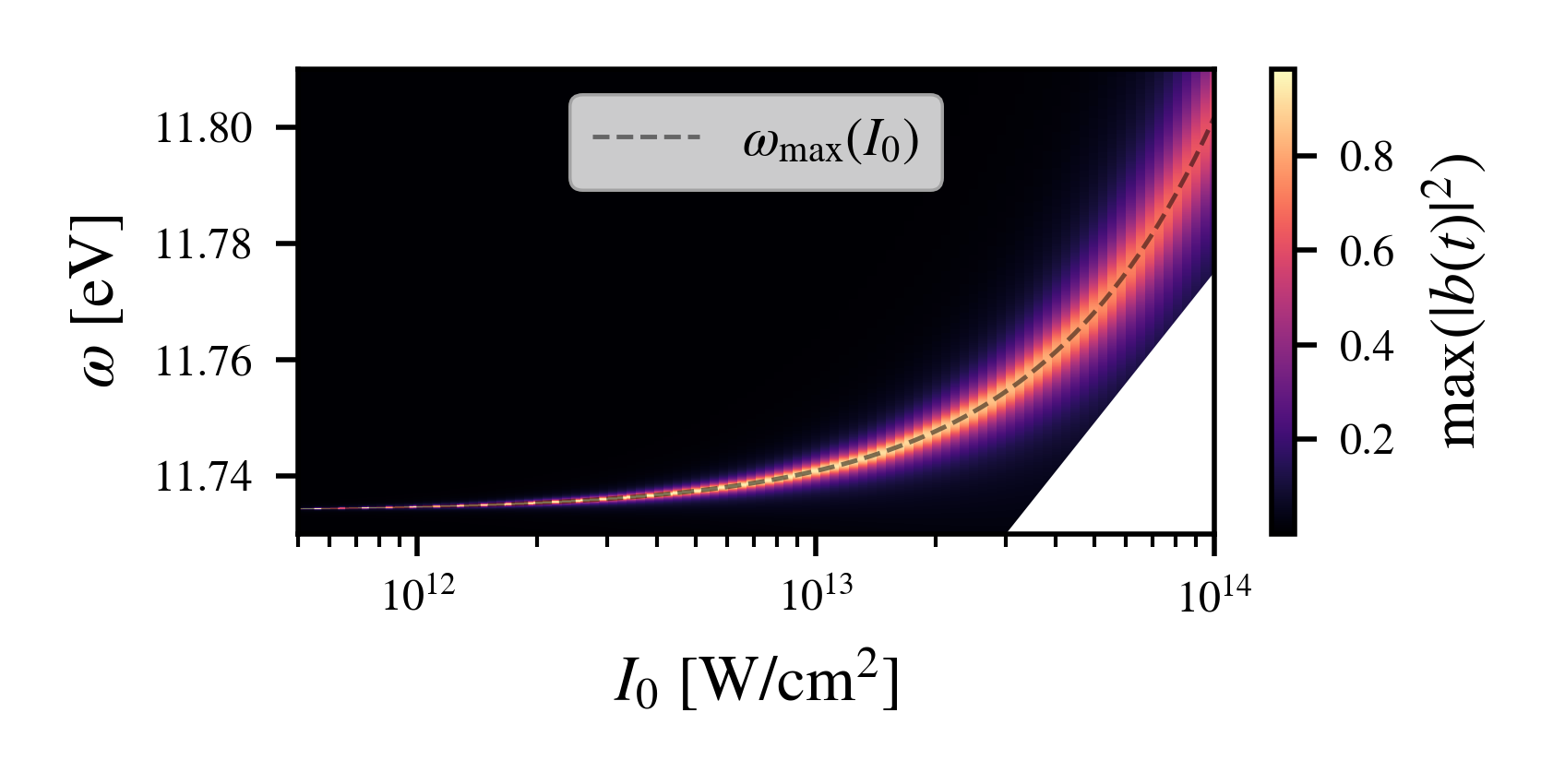}
    \caption{Maximum of the excited state population, $\max(|b(t)|^2)$, in the parameter space of $\omega$ and $I_0$ (1s$^2$-1s3d system). The white space signifies an off-resonant high-intensity region of the parameter space, in which the effective two-level Hamiltonian does not apply.}
    \label{fig:resonancecondition}
\end{figure}

\textit{Optimization of 2-PRO}---In Fig.~\ref{fig:poprabi}, we showed results for the resonantly driven two-photon transition in (a) hydrogen (1s $\leftrightarrow$ 2s) and helium: (b) 1s$^2$ $\leftrightarrow$ 1s2s, (c) 1s$^2$ $\leftrightarrow$ 1s3d, for rectangular pulses. The results were obtained by interpolating $H_\eff$ between different $\omega$ and $I_0$, and evaluating $H_\eff(\omega, I_0)$ at the dressed-resonance frequency $\omega_{\max}(I_0)$. We find this Stark-shifted resonance by calculating the maximum population of the ES $\max(|b(t)|^2)$ depending on $\omega$ and $I_0$. The results are shown in Fig.~\ref{fig:resonancecondition} for the 1s$^2$-1s3d transition. The resonance condition is extremely narrow at low intensities. With higher intensities, the condition becomes more lenient, allowing for detunings of $\sim\SI{5}{\milli\eV}$ or more. 

\begin{figure*}[t]
\centering
\includegraphics[width=\linewidth]{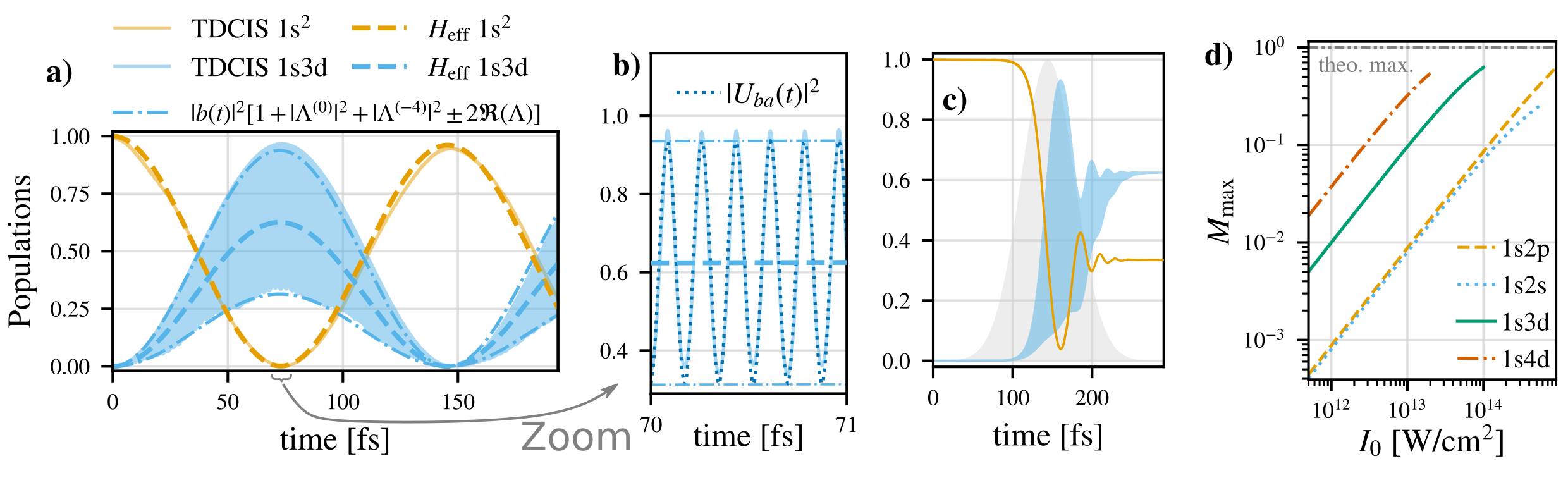}
\caption{(a) 2-PRO from 1s$^2$ to 1s3d, induced by a monochromatic field, from with \textit{ab-initio} (solid) and $H_\eff$ (dashed) calculations. CR oscillations are the blue area. Outlines of the CR oscillations in dashed-dotted. There is no yellow area due to absence of CR oscillations in the GS. (b) Zoom-in on one femtosecond, additionally plotting $|U_{ba}(t)|^2$ from Eq.~\eqref{eq:crpopb2} (dotted). (c) Results for a Gaussian envelope (shaded grey). (d) Maximum span of CR oscillations $M_{\max}$ against intensity for one-photon transition to 1s2p, and two-photon transition to 1s2s, 1s3d, and 1s4d. Theoretical maximum is at $M_{\max} = 1$ (representing CR oscillations that lead to a complete population transfer on sub-cycle scales). See main text for laser parameters in (a) and (c).}
\label{fig:CRoscillations}
\end{figure*}    

\textit{Numerical results}---In Fig.~\ref{fig:CRoscillations} (a) we show evidence for giant CR oscillations in the 1s3d state at $I_0 = \SI{e14}{\watt\per\square\cm}$ and $\omega=\SI{11.8017}{\eV}$ (blue area). The mean populations (dashed) and the analytically calculated outlines of the CR oscillations (dashed-dotted) reproduce the TDCIS calculations accurately. In panel (b), we zoom in on a single femtosecond to show the sub-cycle modulation of the ES on the attosecond timescale. Here, we further plot $|U_{ba}(t)|^2$ from Eq.~\eqref{eq:crpopb2} to show the excellent --  sub-cycle -- agreement of the theoretical model with numerical TDCIS simulations. 

The 1s$^2$-1s3d system displays complete non-reciprocality in the CR oscillations: The GS evolves only with the optical cycle-averaged population $|a(t)|^2$ due to $\ket{a,0}$, whereas the ES is modulated with CR oscillations that are as large as the 2-PRO themselves. The asymmetry stems from the vastly different strengths in which the GS and ES couple to the $\Q$-space, which we exploited in the derivation of Eq.~\eqref{eq:crpopb2}. This especially applies to the coupling of the ES with the nearby Rydberg states: The 1s3p, 1s4p, 1s5p, 1s4f, and 1s5f states have the largest population in $\Q$. Remarkably, despite violent exchange of photons to the $\Q$-space via $\ket{b,-2}$, the state coherently returns to the initial $\ket{a,0}$ in $\P$ after the completed Rabi cycle. 
The electron remains bound, yet is highly polarized, switching between d-character and p-/f-character on an attosecond time scale, following the instantaneous laser field. Unlike how free electrons wiggle in laser fields, the underlying assumption in the strong-field approximation \cite{lewensteinTheoryHighharmonicGeneration1994}, our reported dynamics arise due to simultaneous confinement by the atomic potential and the action of the instantaneous laser field. 

\textit{Discussion}---Following the several dismissals of CR phenomena in the literature \cite{shirleySolutionSchrodingerEquation1965, lagattutaAbovethresholdIonizationAtomic1993, dorrTimeEvolutionTwophoton1997}, it is imperative to motivate that the proposed giant CR effects can be generated with realistic pulse forms and observed experimentally through possible detection schemes. We find that the breaking of CR reciprocality is robust against changes in the envelope of the field, as shown in Fig.~\ref{fig:CRoscillations} (c) using TDCIS. The simulation was performed with a Gaussian intensity envelope, $I(t) = I_0 \e^{-\ln{2} [2t/\tau]^2}$, where $I_0 = \SI{2e14}{\watt\per\square\cm}$ and $\tau=\mathrm{FWHM} = \SI{80}{\fs}$ at a frequency $\omega = \SI{11.86}{\eV}$, corresponding to $\sim \SI{105}{\nano\meter}$. These parameters are reasonable for seeded FEL sources, such as the externally seeded FEL at FERMI \cite{allariaetal.HighlyCoherentStable2012, allariaetal.TwostageSeededSoftXray2013, finettietal.PulseDurationSeeded2017}. 

It is also important to understand the generality of the CR effects. Can they be observed in other UV/XUV-dressed systems? In Fig.~\ref{fig:CRoscillations}~(d), we compare the maximum span of CR oscillations $M_{\max}$ between different experimentally relevant systems. In 1-PRO, 1s$^2$-1s2p, the CR oscillations are weaker for a given intensity, so that intensities as large as $\SI{9e14}{\watt\per\square\cm}$ are required to generate giant CR oscillations. Meanwhile, the two-photon transition 1s$^2$-1s2s is too damped, making it unsuitable for our purposes. 
A possible way to enhance CR oscillations is to target higher Rydberg states that couple to $\mathcal{Q}$-space states with stronger dipole moments. We demonstrate this in panel~(d) by including results for the 1s4d state. Generating giant CR oscillations in higher Rydberg states requires less intense, but longer pulses.

\textit{Experimental observation}---We propose that the giant CR oscillations can be probed experimentally in two complementary ways: in the frequency or time domain. In order to detect the CR oscillations in the frequency domain, we require an auxiliary probe field that can ionize the atom from its excited state. 
The probe frequency should not be a harmonic of the driving resonant dressing field, $\omega_\mathrm{probe} \ne m\omega$. 
The signature of giant CR oscillations will then be observed at photoelectron kinetic energies close to $E_\mathrm{kin}\approx 4\omega\pm\omega_\mathrm{probe}-E_\mathrm{bin}>0$, where $E_\mathrm{bin}$ is the binding energy of the atom.  

The time-dependence of the CR oscillations can be probed by a synchronized auxiliary attosecond pulse. Prior work with seeded FEL have shown that it is possible to generate {\it attosecond waveforms} \cite{marojuetal.AttosecondPulseShaping2020} that open the possibility to temporally resolve the CR oscillations by photoelectron spectroscopy. 

\textit{Conclusion}---We have demonstrated via \textit{ab initio} simulations that giant CR effects can be induced with intense UV/XUV fields in helium on attosecond timescales. 
This opens the gates to an unexplored perturbative ultrastrong coupling regime. For semi-classical systems, this implies a dependence on the instantaneous electric field. Contrary to the simple two-level system, where CR oscillations are reciprocal, we find that only the ES is modulated on a sub-cycle scale, while the GS is effectively shielded from the $\Q$-space. This effect is remarkably strong in 2-PRO from 1s$^2$ to 1s3d. 
We have explored these effects quantitatively through a theoretical approach for describing CR oscillations, enabled by effective Hamiltonian theory and Floquet theory. 
Given the power of Floquet approaches in different areas of physics \cite{chuFloquetTheoremGeneralized2004, rudnerBandStructureEngineering2020, okaFloquetEngineeringQuantum2019}, we expect that our approach will be fruitful in a range of fields.

\textit{Acknowledgements}---We acknowledge Yijie Liao for useful discussions. JMD acknowledges support from the Olle Engkvist Foundation: 194-0734 and the Knut and Alice Wallenberg Foundation: 2019.0154 \& 2024.0212 and the Swedish Research Council: 2024-04247.

\bibliography{references}

\begin{thebibliography}{68}%
\makeatletter
\providecommand \@ifxundefined [1]{%
 \@ifx{#1\undefined}
}%
\providecommand \@ifnum [1]{%
 \ifnum #1\expandafter \@firstoftwo
 \else \expandafter \@secondoftwo
 \fi
}%
\providecommand \@ifx [1]{%
 \ifx #1\expandafter \@firstoftwo
 \else \expandafter \@secondoftwo
 \fi
}%
\providecommand \natexlab [1]{#1}%
\providecommand \enquote  [1]{``#1''}%
\providecommand \bibnamefont  [1]{#1}%
\providecommand \bibfnamefont [1]{#1}%
\providecommand \citenamefont [1]{#1}%
\providecommand \href@noop [0]{\@secondoftwo}%
\providecommand \href [0]{\begingroup \@sanitize@url \@href}%
\providecommand \@href[1]{\@@startlink{#1}\@@href}%
\providecommand \@@href[1]{\endgroup#1\@@endlink}%
\providecommand \@sanitize@url [0]{\catcode `\\12\catcode `\$12\catcode `\&12\catcode `\#12\catcode `\^12\catcode `\_12\catcode `\%12\relax}%
\providecommand \@@startlink[1]{}%
\providecommand \@@endlink[0]{}%
\providecommand \url  [0]{\begingroup\@sanitize@url \@url }%
\providecommand \@url [1]{\endgroup\@href {#1}{\urlprefix }}%
\providecommand \urlprefix  [0]{URL }%
\providecommand \Eprint [0]{\href }%
\providecommand \doibase [0]{https://doi.org/}%
\providecommand \selectlanguage [0]{\@gobble}%
\providecommand \bibinfo  [0]{\@secondoftwo}%
\providecommand \bibfield  [0]{\@secondoftwo}%
\providecommand \translation [1]{[#1]}%
\providecommand \BibitemOpen [0]{}%
\providecommand \bibitemStop [0]{}%
\providecommand \bibitemNoStop [0]{.\EOS\space}%
\providecommand \EOS [0]{\spacefactor3000\relax}%
\providecommand \BibitemShut  [1]{\csname bibitem#1\endcsname}%
\let\auto@bib@innerbib\@empty
\bibitem [{\citenamefont {{Cohen-Tannoudji}}\ \emph {et~al.}(1998)\citenamefont {{Cohen-Tannoudji}}, \citenamefont {Grynberg},\ and\ \citenamefont {{Dupont-Roc}}}]{cohen-tannoudjiAtomPhotonInteractionsBasic1998}%
  \BibitemOpen
  \bibfield  {author} {\bibinfo {author} {\bibfnamefont {C.}~\bibnamefont {{Cohen-Tannoudji}}}, \bibinfo {author} {\bibfnamefont {G.}~\bibnamefont {Grynberg}},\ and\ \bibinfo {author} {\bibfnamefont {J.}~\bibnamefont {{Dupont-Roc}}},\ }\href@noop {} {\emph {\bibinfo {title} {Atom-{{Photon Interactions}}: {{Basic Processes}} and {{Applications}}}}}\ (\bibinfo  {publisher} {Wiley-VCH},\ \bibinfo {address} {New York},\ \bibinfo {year} {1998})\BibitemShut {NoStop}%
\bibitem [{\citenamefont {Krausz}\ and\ \citenamefont {Ivanov}(2009)}]{krauszAttosecondPhysics2009}%
  \BibitemOpen
  \bibfield  {author} {\bibinfo {author} {\bibfnamefont {F.}~\bibnamefont {Krausz}}\ and\ \bibinfo {author} {\bibfnamefont {M.}~\bibnamefont {Ivanov}},\ }\bibfield  {title} {\bibinfo {title} {Attosecond physics},\ }\href {https://doi.org/10.1103/RevModPhys.81.163} {\bibfield  {journal} {\bibinfo  {journal} {Reviews of Modern Physics}\ }\textbf {\bibinfo {volume} {81}},\ \bibinfo {pages} {163} (\bibinfo {year} {2009})}\BibitemShut {NoStop}%
\bibitem [{\citenamefont {Frisk~Kockum}\ \emph {et~al.}(2019)\citenamefont {Frisk~Kockum}, \citenamefont {Miranowicz}, \citenamefont {De~Liberato}, \citenamefont {Savasta},\ and\ \citenamefont {Nori}}]{friskkockumUltrastrongCouplingLight2019}%
  \BibitemOpen
  \bibfield  {author} {\bibinfo {author} {\bibfnamefont {A.}~\bibnamefont {Frisk~Kockum}}, \bibinfo {author} {\bibfnamefont {A.}~\bibnamefont {Miranowicz}}, \bibinfo {author} {\bibfnamefont {S.}~\bibnamefont {De~Liberato}}, \bibinfo {author} {\bibfnamefont {S.}~\bibnamefont {Savasta}},\ and\ \bibinfo {author} {\bibfnamefont {F.}~\bibnamefont {Nori}},\ }\bibfield  {title} {\bibinfo {title} {Ultrastrong coupling between light and matter},\ }\href {https://doi.org/10.1038/s42254-018-0006-2} {\bibfield  {journal} {\bibinfo  {journal} {Nature Reviews Physics}\ }\textbf {\bibinfo {volume} {1}},\ \bibinfo {pages} {19} (\bibinfo {year} {2019})}\BibitemShut {NoStop}%
\bibitem [{\citenamefont {Bloch}\ and\ \citenamefont {Siegert}(1940)}]{blochMagneticResonanceNonrotating1940}%
  \BibitemOpen
  \bibfield  {author} {\bibinfo {author} {\bibfnamefont {F.}~\bibnamefont {Bloch}}\ and\ \bibinfo {author} {\bibfnamefont {A.}~\bibnamefont {Siegert}},\ }\bibfield  {title} {\bibinfo {title} {Magnetic {{Resonance}} for {{Nonrotating Fields}}},\ }\href {https://doi.org/10.1103/PhysRev.57.522} {\bibfield  {journal} {\bibinfo  {journal} {Physical Review}\ }\textbf {\bibinfo {volume} {57}},\ \bibinfo {pages} {522} (\bibinfo {year} {1940})}\BibitemShut {NoStop}%
\bibitem [{\citenamefont {Autler}\ and\ \citenamefont {Townes}(1955)}]{autlerStarkEffectRapidly1955}%
  \BibitemOpen
  \bibfield  {author} {\bibinfo {author} {\bibfnamefont {S.~H.}\ \bibnamefont {Autler}}\ and\ \bibinfo {author} {\bibfnamefont {C.~H.}\ \bibnamefont {Townes}},\ }\bibfield  {title} {\bibinfo {title} {Stark {{Effect}} in {{Rapidly Varying Fields}}},\ }\href {https://doi.org/10.1103/PhysRev.100.703} {\bibfield  {journal} {\bibinfo  {journal} {Physical Review}\ }\textbf {\bibinfo {volume} {100}},\ \bibinfo {pages} {703} (\bibinfo {year} {1955})}\BibitemShut {NoStop}%
\bibitem [{\citenamefont {Shirley}(1965)}]{shirleySolutionSchrodingerEquation1965}%
  \BibitemOpen
  \bibfield  {author} {\bibinfo {author} {\bibfnamefont {J.~H.}\ \bibnamefont {Shirley}},\ }\bibfield  {title} {\bibinfo {title} {Solution of the {{Schr{\"o}dinger Equation}} with a {{Hamiltonian Periodic}} in {{Time}}},\ }\href {https://doi.org/10.1103/PhysRev.138.B979} {\bibfield  {journal} {\bibinfo  {journal} {Physical Review}\ }\textbf {\bibinfo {volume} {138}},\ \bibinfo {pages} {B979} (\bibinfo {year} {1965})}\BibitemShut {NoStop}%
\bibitem [{\citenamefont {LaGattuta}(1993)}]{lagattutaAbovethresholdIonizationAtomic1993}%
  \BibitemOpen
  \bibfield  {author} {\bibinfo {author} {\bibfnamefont {K.~J.}\ \bibnamefont {LaGattuta}},\ }\bibfield  {title} {\bibinfo {title} {Above-threshold ionization of atomic hydrogen via resonant intermediate states},\ }\href {https://doi.org/10.1103/PhysRevA.47.1560} {\bibfield  {journal} {\bibinfo  {journal} {Physical Review A}\ }\textbf {\bibinfo {volume} {47}},\ \bibinfo {pages} {1560} (\bibinfo {year} {1993})}\BibitemShut {NoStop}%
\bibitem [{\citenamefont {D{\"o}rr}\ \emph {et~al.}(1997)\citenamefont {D{\"o}rr}, \citenamefont {Latinne},\ and\ \citenamefont {Joachain}}]{dorrTimeEvolutionTwophoton1997}%
  \BibitemOpen
  \bibfield  {author} {\bibinfo {author} {\bibfnamefont {M.}~\bibnamefont {D{\"o}rr}}, \bibinfo {author} {\bibfnamefont {O.}~\bibnamefont {Latinne}},\ and\ \bibinfo {author} {\bibfnamefont {C.~J.}\ \bibnamefont {Joachain}},\ }\bibfield  {title} {\bibinfo {title} {Time evolution of two-photon population transfer between the 1s and 2s states of a hydrogen atom},\ }\href {https://doi.org/10.1103/PhysRevA.55.3697} {\bibfield  {journal} {\bibinfo  {journal} {Physical Review A}\ }\textbf {\bibinfo {volume} {55}},\ \bibinfo {pages} {3697} (\bibinfo {year} {1997})}\BibitemShut {NoStop}%
\bibitem [{\citenamefont {Kaiser}\ \emph {et~al.}(2013)\citenamefont {Kaiser}, \citenamefont {Brand}, \citenamefont {Gl{\"a}ssl}, \citenamefont {Vagov}, \citenamefont {Axt},\ and\ \citenamefont {Pietsch}}]{kaiserPhotoionizationResonantlyDriven2013}%
  \BibitemOpen
  \bibfield  {author} {\bibinfo {author} {\bibfnamefont {B.}~\bibnamefont {Kaiser}}, \bibinfo {author} {\bibfnamefont {A.}~\bibnamefont {Brand}}, \bibinfo {author} {\bibfnamefont {M.}~\bibnamefont {Gl{\"a}ssl}}, \bibinfo {author} {\bibfnamefont {A.}~\bibnamefont {Vagov}}, \bibinfo {author} {\bibfnamefont {V.~M.}\ \bibnamefont {Axt}},\ and\ \bibinfo {author} {\bibfnamefont {U.}~\bibnamefont {Pietsch}},\ }\bibfield  {title} {\bibinfo {title} {Photoionization of resonantly driven atomic states by an extreme ultraviolet-free-electron laser: Intensity dependence and renormalization of {{Rabi}} frequencies},\ }\href {https://doi.org/10.1088/1367-2630/15/9/093016} {\bibfield  {journal} {\bibinfo  {journal} {New Journal of Physics}\ }\textbf {\bibinfo {volume} {15}},\ \bibinfo {pages} {093016} (\bibinfo {year} {2013})}\BibitemShut {NoStop}%
\bibitem [{\citenamefont {Xie}\ \emph {et~al.}(2017)\citenamefont {Xie}, \citenamefont {Zhong}, \citenamefont {Batchelor},\ and\ \citenamefont {Lee}}]{xieQuantumRabiModel2017}%
  \BibitemOpen
  \bibfield  {author} {\bibinfo {author} {\bibfnamefont {Q.}~\bibnamefont {Xie}}, \bibinfo {author} {\bibfnamefont {H.}~\bibnamefont {Zhong}}, \bibinfo {author} {\bibfnamefont {M.~T.}\ \bibnamefont {Batchelor}},\ and\ \bibinfo {author} {\bibfnamefont {C.}~\bibnamefont {Lee}},\ }\bibfield  {title} {\bibinfo {title} {The quantum {{Rabi}} model: Solution and dynamics},\ }\href {https://doi.org/10.1088/1751-8121/aa5a65} {\bibfield  {journal} {\bibinfo  {journal} {Journal of Physics A: Mathematical and Theoretical}\ }\textbf {\bibinfo {volume} {50}},\ \bibinfo {pages} {113001} (\bibinfo {year} {2017})}\BibitemShut {NoStop}%
\bibitem [{\citenamefont {Jaynes}\ and\ \citenamefont {Cummings}(1963)}]{jaynesComparisonQuantumSemiclassical1963}%
  \BibitemOpen
  \bibfield  {author} {\bibinfo {author} {\bibfnamefont {E.}~\bibnamefont {Jaynes}}\ and\ \bibinfo {author} {\bibfnamefont {F.}~\bibnamefont {Cummings}},\ }\bibfield  {title} {\bibinfo {title} {Comparison of quantum and semiclassical radiation theories with application to the beam maser},\ }\href {https://doi.org/10.1109/PROC.1963.1664} {\bibfield  {journal} {\bibinfo  {journal} {Proceedings of the IEEE}\ }\textbf {\bibinfo {volume} {51}},\ \bibinfo {pages} {89} (\bibinfo {year} {1963})}\BibitemShut {NoStop}%
\bibitem [{\citenamefont {{Forn-D{\'i}az}}\ \emph {et~al.}(2019)\citenamefont {{Forn-D{\'i}az}}, \citenamefont {Lamata}, \citenamefont {Rico}, \citenamefont {Kono},\ and\ \citenamefont {Solano}}]{forn-diazUltrastrongCouplingRegimes2019}%
  \BibitemOpen
  \bibfield  {author} {\bibinfo {author} {\bibfnamefont {P.}~\bibnamefont {{Forn-D{\'i}az}}}, \bibinfo {author} {\bibfnamefont {L.}~\bibnamefont {Lamata}}, \bibinfo {author} {\bibfnamefont {E.}~\bibnamefont {Rico}}, \bibinfo {author} {\bibfnamefont {J.}~\bibnamefont {Kono}},\ and\ \bibinfo {author} {\bibfnamefont {E.}~\bibnamefont {Solano}},\ }\bibfield  {title} {\bibinfo {title} {Ultrastrong coupling regimes of light-matter interaction},\ }\href {https://doi.org/10.1103/RevModPhys.91.025005} {\bibfield  {journal} {\bibinfo  {journal} {Reviews of Modern Physics}\ }\textbf {\bibinfo {volume} {91}},\ \bibinfo {pages} {025005} (\bibinfo {year} {2019})}\BibitemShut {NoStop}%
\bibitem [{\citenamefont {Rossatto}\ \emph {et~al.}(2017)\citenamefont {Rossatto}, \citenamefont {{Villas-B{\^o}as}}, \citenamefont {Sanz},\ and\ \citenamefont {Solano}}]{rossattoSpectralClassificationCoupling2017}%
  \BibitemOpen
  \bibfield  {author} {\bibinfo {author} {\bibfnamefont {D.~Z.}\ \bibnamefont {Rossatto}}, \bibinfo {author} {\bibfnamefont {C.~J.}\ \bibnamefont {{Villas-B{\^o}as}}}, \bibinfo {author} {\bibfnamefont {M.}~\bibnamefont {Sanz}},\ and\ \bibinfo {author} {\bibfnamefont {E.}~\bibnamefont {Solano}},\ }\bibfield  {title} {\bibinfo {title} {Spectral classification of coupling regimes in the quantum {{Rabi}} model},\ }\href {https://doi.org/10.1103/PhysRevA.96.013849} {\bibfield  {journal} {\bibinfo  {journal} {Physical Review A}\ }\textbf {\bibinfo {volume} {96}},\ \bibinfo {pages} {013849} (\bibinfo {year} {2017})}\BibitemShut {NoStop}%
\bibitem [{\citenamefont {Lewenstein}\ \emph {et~al.}(1994)\citenamefont {Lewenstein}, \citenamefont {Balcou}, \citenamefont {Ivanov}, \citenamefont {L'Huillier},\ and\ \citenamefont {Corkum}}]{lewensteinTheoryHighharmonicGeneration1994}%
  \BibitemOpen
  \bibfield  {author} {\bibinfo {author} {\bibfnamefont {M.}~\bibnamefont {Lewenstein}}, \bibinfo {author} {\bibfnamefont {{\relax Ph}.}~\bibnamefont {Balcou}}, \bibinfo {author} {\bibfnamefont {M.~{\relax Yu}.}\ \bibnamefont {Ivanov}}, \bibinfo {author} {\bibfnamefont {A.}~\bibnamefont {L'Huillier}},\ and\ \bibinfo {author} {\bibfnamefont {P.~B.}\ \bibnamefont {Corkum}},\ }\bibfield  {title} {\bibinfo {title} {Theory of high-harmonic generation by low-frequency laser fields},\ }\href {https://doi.org/10.1103/PhysRevA.49.2117} {\bibfield  {journal} {\bibinfo  {journal} {Physical Review A}\ }\textbf {\bibinfo {volume} {49}},\ \bibinfo {pages} {2117} (\bibinfo {year} {1994})}\BibitemShut {NoStop}%
\bibitem [{\citenamefont {Paul}\ \emph {et~al.}(2001)\citenamefont {Paul}, \citenamefont {Toma}, \citenamefont {Breger}, \citenamefont {Mullot}, \citenamefont {Aug{\'e}}, \citenamefont {Balcou}, \citenamefont {Muller},\ and\ \citenamefont {Agostini}}]{paulObservationTrainAttosecond2001}%
  \BibitemOpen
  \bibfield  {author} {\bibinfo {author} {\bibfnamefont {P.~M.}\ \bibnamefont {Paul}}, \bibinfo {author} {\bibfnamefont {E.~S.}\ \bibnamefont {Toma}}, \bibinfo {author} {\bibfnamefont {P.}~\bibnamefont {Breger}}, \bibinfo {author} {\bibfnamefont {G.}~\bibnamefont {Mullot}}, \bibinfo {author} {\bibfnamefont {F.}~\bibnamefont {Aug{\'e}}}, \bibinfo {author} {\bibfnamefont {{\relax Ph}.}~\bibnamefont {Balcou}}, \bibinfo {author} {\bibfnamefont {H.~G.}\ \bibnamefont {Muller}},\ and\ \bibinfo {author} {\bibfnamefont {P.}~\bibnamefont {Agostini}},\ }\bibfield  {title} {\bibinfo {title} {Observation of a {{Train}} of {{Attosecond Pulses}} from {{High Harmonic Generation}}},\ }\href {https://doi.org/10.1126/science.1059413} {\bibfield  {journal} {\bibinfo  {journal} {Science}\ }\textbf {\bibinfo {volume} {292}},\ \bibinfo {pages} {1689} (\bibinfo {year} {2001})}\BibitemShut {NoStop}%
\bibitem [{\citenamefont {Dahlstr{\"o}m}\ \emph {et~al.}(2012)\citenamefont {Dahlstr{\"o}m}, \citenamefont {L'Huillier},\ and\ \citenamefont {Maquet}}]{dahlstromIntroductionAttosecondDelays2012}%
  \BibitemOpen
  \bibfield  {author} {\bibinfo {author} {\bibfnamefont {J.~M.}\ \bibnamefont {Dahlstr{\"o}m}}, \bibinfo {author} {\bibfnamefont {A.}~\bibnamefont {L'Huillier}},\ and\ \bibinfo {author} {\bibfnamefont {A.}~\bibnamefont {Maquet}},\ }\bibfield  {title} {\bibinfo {title} {Introduction to attosecond delays in photoionization},\ }\href {https://doi.org/10.1088/0953-4075/45/18/183001} {\bibfield  {journal} {\bibinfo  {journal} {Journal of Physics B: Atomic, Molecular and Optical Physics}\ }\textbf {\bibinfo {volume} {45}},\ \bibinfo {pages} {183001} (\bibinfo {year} {2012})}\BibitemShut {NoStop}%
\bibitem [{\citenamefont {Calegari}\ \emph {et~al.}(2016)\citenamefont {Calegari}, \citenamefont {Trabattoni}, \citenamefont {Palacios}, \citenamefont {Ayuso}, \citenamefont {Castrovilli}, \citenamefont {Greenwood}, \citenamefont {Decleva}, \citenamefont {Mart{\'i}n},\ and\ \citenamefont {Nisoli}}]{calegariChargeMigrationInduced2016}%
  \BibitemOpen
  \bibfield  {author} {\bibinfo {author} {\bibfnamefont {F.}~\bibnamefont {Calegari}}, \bibinfo {author} {\bibfnamefont {A.}~\bibnamefont {Trabattoni}}, \bibinfo {author} {\bibfnamefont {A.}~\bibnamefont {Palacios}}, \bibinfo {author} {\bibfnamefont {D.}~\bibnamefont {Ayuso}}, \bibinfo {author} {\bibfnamefont {M.~C.}\ \bibnamefont {Castrovilli}}, \bibinfo {author} {\bibfnamefont {J.~B.}\ \bibnamefont {Greenwood}}, \bibinfo {author} {\bibfnamefont {P.}~\bibnamefont {Decleva}}, \bibinfo {author} {\bibfnamefont {F.}~\bibnamefont {Mart{\'i}n}},\ and\ \bibinfo {author} {\bibfnamefont {M.}~\bibnamefont {Nisoli}},\ }\bibfield  {title} {\bibinfo {title} {Charge migration induced by attosecond pulses in bio-relevant molecules},\ }\href {https://doi.org/10.1088/0953-4075/49/14/142001} {\bibfield  {journal} {\bibinfo  {journal} {Journal of Physics B: Atomic, Molecular and Optical Physics}\ }\textbf {\bibinfo {volume} {49}},\ \bibinfo {pages} {142001} (\bibinfo {year} {2016})}\BibitemShut {NoStop}%
\bibitem [{\citenamefont {Vampa}\ \emph {et~al.}(2014)\citenamefont {Vampa}, \citenamefont {McDonald}, \citenamefont {Orlando}, \citenamefont {Klug}, \citenamefont {Corkum},\ and\ \citenamefont {Brabec}}]{vampaTheoreticalAnalysisHighHarmonic2014}%
  \BibitemOpen
  \bibfield  {author} {\bibinfo {author} {\bibfnamefont {G.}~\bibnamefont {Vampa}}, \bibinfo {author} {\bibfnamefont {C.~R.}\ \bibnamefont {McDonald}}, \bibinfo {author} {\bibfnamefont {G.}~\bibnamefont {Orlando}}, \bibinfo {author} {\bibfnamefont {D.~D.}\ \bibnamefont {Klug}}, \bibinfo {author} {\bibfnamefont {P.~B.}\ \bibnamefont {Corkum}},\ and\ \bibinfo {author} {\bibfnamefont {T.}~\bibnamefont {Brabec}},\ }\bibfield  {title} {\bibinfo {title} {Theoretical {{Analysis}} of {{High-Harmonic Generation}} in {{Solids}}},\ }\href {https://doi.org/10.1103/PhysRevLett.113.073901} {\bibfield  {journal} {\bibinfo  {journal} {Physical Review Letters}\ }\textbf {\bibinfo {volume} {113}},\ \bibinfo {pages} {073901} (\bibinfo {year} {2014})}\BibitemShut {NoStop}%
\bibitem [{\citenamefont {Yue}\ and\ \citenamefont {Gaarde}(2022)}]{yueIntroductionTheoryHighharmonic2022}%
  \BibitemOpen
  \bibfield  {author} {\bibinfo {author} {\bibfnamefont {L.}~\bibnamefont {Yue}}\ and\ \bibinfo {author} {\bibfnamefont {M.~B.}\ \bibnamefont {Gaarde}},\ }\bibfield  {title} {\bibinfo {title} {Introduction to theory of high-harmonic generation in solids: Tutorial},\ }\href {https://doi.org/10.1364/JOSAB.448602} {\bibfield  {journal} {\bibinfo  {journal} {JOSA B}\ }\textbf {\bibinfo {volume} {39}},\ \bibinfo {pages} {535} (\bibinfo {year} {2022})}\BibitemShut {NoStop}%
\bibitem [{\citenamefont {Pazourek}\ \emph {et~al.}(2015)\citenamefont {Pazourek}, \citenamefont {Nagele},\ and\ \citenamefont {Burgd{\"o}rfer}}]{pazourekAttosecondChronoscopyPhotoemission2015}%
  \BibitemOpen
  \bibfield  {author} {\bibinfo {author} {\bibfnamefont {R.}~\bibnamefont {Pazourek}}, \bibinfo {author} {\bibfnamefont {S.}~\bibnamefont {Nagele}},\ and\ \bibinfo {author} {\bibfnamefont {J.}~\bibnamefont {Burgd{\"o}rfer}},\ }\bibfield  {title} {\bibinfo {title} {Attosecond chronoscopy of photoemission},\ }\href {https://doi.org/10.1103/RevModPhys.87.765} {\bibfield  {journal} {\bibinfo  {journal} {Reviews of Modern Physics}\ }\textbf {\bibinfo {volume} {87}},\ \bibinfo {pages} {765} (\bibinfo {year} {2015})}\BibitemShut {NoStop}%
\bibitem [{\citenamefont {Chini}\ \emph {et~al.}(2012)\citenamefont {Chini}, \citenamefont {Zhao}, \citenamefont {Wang}, \citenamefont {Cheng}, \citenamefont {Hu},\ and\ \citenamefont {Chang}}]{chiniSubcycleAcStark2012}%
  \BibitemOpen
  \bibfield  {author} {\bibinfo {author} {\bibfnamefont {M.}~\bibnamefont {Chini}}, \bibinfo {author} {\bibfnamefont {B.}~\bibnamefont {Zhao}}, \bibinfo {author} {\bibfnamefont {H.}~\bibnamefont {Wang}}, \bibinfo {author} {\bibfnamefont {Y.}~\bibnamefont {Cheng}}, \bibinfo {author} {\bibfnamefont {S.~X.}\ \bibnamefont {Hu}},\ and\ \bibinfo {author} {\bibfnamefont {Z.}~\bibnamefont {Chang}},\ }\bibfield  {title} {\bibinfo {title} {Subcycle ac {{Stark Shift}} of {{Helium Excited States Probed}} with {{Isolated Attosecond Pulses}}},\ }\href {https://doi.org/10.1103/PhysRevLett.109.073601} {\bibfield  {journal} {\bibinfo  {journal} {Physical Review Letters}\ }\textbf {\bibinfo {volume} {109}},\ \bibinfo {pages} {073601} (\bibinfo {year} {2012})}\BibitemShut {NoStop}%
\bibitem [{\citenamefont {Reduzzi}\ \emph {et~al.}(2015)\citenamefont {Reduzzi}, \citenamefont {Hummert}, \citenamefont {Dubrouil}, \citenamefont {Calegari}, \citenamefont {Nisoli}, \citenamefont {Frassetto}, \citenamefont {Poletto}, \citenamefont {Chen}, \citenamefont {Wu}, \citenamefont {Gaarde}, \citenamefont {Schafer},\ and\ \citenamefont {Sansone}}]{reduzziPolarizationControlAbsorption2015}%
  \BibitemOpen
  \bibfield  {author} {\bibinfo {author} {\bibfnamefont {M.}~\bibnamefont {Reduzzi}}, \bibinfo {author} {\bibfnamefont {J.}~\bibnamefont {Hummert}}, \bibinfo {author} {\bibfnamefont {A.}~\bibnamefont {Dubrouil}}, \bibinfo {author} {\bibfnamefont {F.}~\bibnamefont {Calegari}}, \bibinfo {author} {\bibfnamefont {M.}~\bibnamefont {Nisoli}}, \bibinfo {author} {\bibfnamefont {F.}~\bibnamefont {Frassetto}}, \bibinfo {author} {\bibfnamefont {L.}~\bibnamefont {Poletto}}, \bibinfo {author} {\bibfnamefont {S.}~\bibnamefont {Chen}}, \bibinfo {author} {\bibfnamefont {M.}~\bibnamefont {Wu}}, \bibinfo {author} {\bibfnamefont {M.~B.}\ \bibnamefont {Gaarde}}, \bibinfo {author} {\bibfnamefont {K.}~\bibnamefont {Schafer}},\ and\ \bibinfo {author} {\bibfnamefont {G.}~\bibnamefont {Sansone}},\ }\bibfield  {title} {\bibinfo {title} {Polarization control of absorption of virtual dressed states in helium},\ }\href {https://doi.org/10.1103/PhysRevA.92.033408} {\bibfield  {journal} {\bibinfo  {journal} {Physical Review A}\ }\textbf
  {\bibinfo {volume} {92}},\ \bibinfo {pages} {033408} (\bibinfo {year} {2015})}\BibitemShut {NoStop}%
\bibitem [{\citenamefont {Anand}\ \emph {et~al.}(2017)\citenamefont {Anand}, \citenamefont {Pabst}, \citenamefont {Kwon},\ and\ \citenamefont {Kim}}]{anandAttosecondCounterrotatingwaveEffect2017}%
  \BibitemOpen
  \bibfield  {author} {\bibinfo {author} {\bibfnamefont {M.}~\bibnamefont {Anand}}, \bibinfo {author} {\bibfnamefont {S.}~\bibnamefont {Pabst}}, \bibinfo {author} {\bibfnamefont {O.}~\bibnamefont {Kwon}},\ and\ \bibinfo {author} {\bibfnamefont {D.~E.}\ \bibnamefont {Kim}},\ }\bibfield  {title} {\bibinfo {title} {Attosecond counter-rotating-wave effect in xenon driven by strong fields},\ }\href {https://doi.org/10.1103/PhysRevA.95.053420} {\bibfield  {journal} {\bibinfo  {journal} {Physical Review A}\ }\textbf {\bibinfo {volume} {95}},\ \bibinfo {pages} {053420} (\bibinfo {year} {2017})}\BibitemShut {NoStop}%
\bibitem [{\citenamefont {Stammer}\ \emph {et~al.}(2023)\citenamefont {Stammer}, \citenamefont {{Rivera-Dean}}, \citenamefont {Maxwell}, \citenamefont {Lamprou}, \citenamefont {Ord{\'o}{\~n}ez}, \citenamefont {Ciappina}, \citenamefont {Tzallas},\ and\ \citenamefont {Lewenstein}}]{stammerQuantumElectrodynamicsIntense2023}%
  \BibitemOpen
  \bibfield  {author} {\bibinfo {author} {\bibfnamefont {P.}~\bibnamefont {Stammer}}, \bibinfo {author} {\bibfnamefont {J.}~\bibnamefont {{Rivera-Dean}}}, \bibinfo {author} {\bibfnamefont {A.}~\bibnamefont {Maxwell}}, \bibinfo {author} {\bibfnamefont {T.}~\bibnamefont {Lamprou}}, \bibinfo {author} {\bibfnamefont {A.}~\bibnamefont {Ord{\'o}{\~n}ez}}, \bibinfo {author} {\bibfnamefont {M.~F.}\ \bibnamefont {Ciappina}}, \bibinfo {author} {\bibfnamefont {P.}~\bibnamefont {Tzallas}},\ and\ \bibinfo {author} {\bibfnamefont {M.}~\bibnamefont {Lewenstein}},\ }\bibfield  {title} {\bibinfo {title} {Quantum {{Electrodynamics}} of {{Intense Laser-Matter Interactions}}: {{A Tool}} for {{Quantum State Engineering}}},\ }\href {https://doi.org/10.1103/PRXQuantum.4.010201} {\bibfield  {journal} {\bibinfo  {journal} {PRX Quantum}\ }\textbf {\bibinfo {volume} {4}},\ \bibinfo {pages} {010201} (\bibinfo {year} {2023})}\BibitemShut {NoStop}%
\bibitem [{\citenamefont {Lewenstein}\ \emph {et~al.}(2021)\citenamefont {Lewenstein}, \citenamefont {Ciappina}, \citenamefont {Pisanty}, \citenamefont {{Rivera-Dean}}, \citenamefont {Stammer}, \citenamefont {Lamprou},\ and\ \citenamefont {Tzallas}}]{lewensteinGenerationOpticalSchrodinger2021}%
  \BibitemOpen
  \bibfield  {author} {\bibinfo {author} {\bibfnamefont {M.}~\bibnamefont {Lewenstein}}, \bibinfo {author} {\bibfnamefont {M.~F.}\ \bibnamefont {Ciappina}}, \bibinfo {author} {\bibfnamefont {E.}~\bibnamefont {Pisanty}}, \bibinfo {author} {\bibfnamefont {J.}~\bibnamefont {{Rivera-Dean}}}, \bibinfo {author} {\bibfnamefont {P.}~\bibnamefont {Stammer}}, \bibinfo {author} {\bibfnamefont {T.}~\bibnamefont {Lamprou}},\ and\ \bibinfo {author} {\bibfnamefont {P.}~\bibnamefont {Tzallas}},\ }\bibfield  {title} {\bibinfo {title} {Generation of optical {{Schr{\"o}dinger}} cat states in intense laser--matter interactions},\ }\href {https://doi.org/10.1038/s41567-021-01317-w} {\bibfield  {journal} {\bibinfo  {journal} {Nature Physics}\ }\textbf {\bibinfo {volume} {17}},\ \bibinfo {pages} {1104} (\bibinfo {year} {2021})}\BibitemShut {NoStop}%
\bibitem [{\citenamefont {Tsatrafyllis}\ \emph {et~al.}(2017)\citenamefont {Tsatrafyllis}, \citenamefont {Kominis}, \citenamefont {Gonoskov},\ and\ \citenamefont {Tzallas}}]{tsatrafyllisHighorderHarmonicsMeasured2017}%
  \BibitemOpen
  \bibfield  {author} {\bibinfo {author} {\bibfnamefont {N.}~\bibnamefont {Tsatrafyllis}}, \bibinfo {author} {\bibfnamefont {I.~K.}\ \bibnamefont {Kominis}}, \bibinfo {author} {\bibfnamefont {I.~A.}\ \bibnamefont {Gonoskov}},\ and\ \bibinfo {author} {\bibfnamefont {P.}~\bibnamefont {Tzallas}},\ }\bibfield  {title} {\bibinfo {title} {High-order harmonics measured by the photon statistics of the infrared driving-field exiting the atomic medium},\ }\href {https://doi.org/10.1038/ncomms15170} {\bibfield  {journal} {\bibinfo  {journal} {Nature Communications}\ }\textbf {\bibinfo {volume} {8}},\ \bibinfo {pages} {15170} (\bibinfo {year} {2017})}\BibitemShut {NoStop}%
\bibitem [{\citenamefont {Gorlach}\ \emph {et~al.}(2020)\citenamefont {Gorlach}, \citenamefont {Neufeld}, \citenamefont {Rivera}, \citenamefont {Cohen},\ and\ \citenamefont {Kaminer}}]{gorlachQuantumopticalNatureHigh2020}%
  \BibitemOpen
  \bibfield  {author} {\bibinfo {author} {\bibfnamefont {A.}~\bibnamefont {Gorlach}}, \bibinfo {author} {\bibfnamefont {O.}~\bibnamefont {Neufeld}}, \bibinfo {author} {\bibfnamefont {N.}~\bibnamefont {Rivera}}, \bibinfo {author} {\bibfnamefont {O.}~\bibnamefont {Cohen}},\ and\ \bibinfo {author} {\bibfnamefont {I.}~\bibnamefont {Kaminer}},\ }\bibfield  {title} {\bibinfo {title} {The quantum-optical nature of high harmonic generation},\ }\href {https://doi.org/10.1038/s41467-020-18218-w} {\bibfield  {journal} {\bibinfo  {journal} {Nature Communications}\ }\textbf {\bibinfo {volume} {11}},\ \bibinfo {pages} {4598} (\bibinfo {year} {2020})}\BibitemShut {NoStop}%
\bibitem [{\citenamefont {Kumlin}\ \emph {et~al.}(2023)\citenamefont {Kumlin}, \citenamefont {Braun}, \citenamefont {Tresp}, \citenamefont {Stiesdal}, \citenamefont {Hofferberth},\ and\ \citenamefont {{Paris-Mandoki}}}]{kumlinQuantumOpticsRydberg2023}%
  \BibitemOpen
  \bibfield  {author} {\bibinfo {author} {\bibfnamefont {J.}~\bibnamefont {Kumlin}}, \bibinfo {author} {\bibfnamefont {C.}~\bibnamefont {Braun}}, \bibinfo {author} {\bibfnamefont {C.}~\bibnamefont {Tresp}}, \bibinfo {author} {\bibfnamefont {N.}~\bibnamefont {Stiesdal}}, \bibinfo {author} {\bibfnamefont {S.}~\bibnamefont {Hofferberth}},\ and\ \bibinfo {author} {\bibfnamefont {A.}~\bibnamefont {{Paris-Mandoki}}},\ }\bibfield  {title} {\bibinfo {title} {Quantum optics with {{Rydberg}} superatoms},\ }\href {https://doi.org/10.1088/2399-6528/acd51d} {\bibfield  {journal} {\bibinfo  {journal} {Journal of Physics Communications}\ }\textbf {\bibinfo {volume} {7}},\ \bibinfo {pages} {052001} (\bibinfo {year} {2023})}\BibitemShut {NoStop}%
\bibitem [{\citenamefont {Beers}\ and\ \citenamefont {Armstrong}(1975)}]{beersExactSolutionRealistic1975}%
  \BibitemOpen
  \bibfield  {author} {\bibinfo {author} {\bibfnamefont {B.~L.}\ \bibnamefont {Beers}}\ and\ \bibinfo {author} {\bibfnamefont {L.}~\bibnamefont {Armstrong}},\ }\bibfield  {title} {\bibinfo {title} {Exact solution of a realistic model for two-photon ionization},\ }\href {https://doi.org/10.1103/PhysRevA.12.2447} {\bibfield  {journal} {\bibinfo  {journal} {Physical Review A}\ }\textbf {\bibinfo {volume} {12}},\ \bibinfo {pages} {2447} (\bibinfo {year} {1975})}\BibitemShut {NoStop}%
\bibitem [{\citenamefont {Holt}\ \emph {et~al.}(1983)\citenamefont {Holt}, \citenamefont {Raymer},\ and\ \citenamefont {Reinhardt}}]{holtTimeDependencesTwo1983}%
  \BibitemOpen
  \bibfield  {author} {\bibinfo {author} {\bibfnamefont {C.~R.}\ \bibnamefont {Holt}}, \bibinfo {author} {\bibfnamefont {M.~G.}\ \bibnamefont {Raymer}},\ and\ \bibinfo {author} {\bibfnamefont {W.~P.}\ \bibnamefont {Reinhardt}},\ }\bibfield  {title} {\bibinfo {title} {Time dependences of two-, three-, and four-photon ionization of atomic hydrogen in the ground 1{\textsuperscript{2}}{{S}} and metastable 2{\textsuperscript{2}}{{S}} states},\ }\href {https://doi.org/10.1103/PhysRevA.27.2971} {\bibfield  {journal} {\bibinfo  {journal} {Physical Review A}\ }\textbf {\bibinfo {volume} {27}},\ \bibinfo {pages} {2971} (\bibinfo {year} {1983})}\BibitemShut {NoStop}%
\bibitem [{\citenamefont {{Emma \emph{et al.}}}(2010)}]{emmaetal.FirstLasingOperation2010}%
  \BibitemOpen
  \bibfield  {author} {\bibinfo {author} {\bibfnamefont {P.}~\bibnamefont {{Emma \emph{et al.}}}},\ }\bibfield  {title} {\bibinfo {title} {First lasing and operation of an {\aa}ngstrom-wavelength free-electron laser},\ }\href {https://doi.org/10.1038/nphoton.2010.176} {\bibfield  {journal} {\bibinfo  {journal} {Nature Photonics}\ }\textbf {\bibinfo {volume} {4}},\ \bibinfo {pages} {641} (\bibinfo {year} {2010})}\BibitemShut {NoStop}%
\bibitem [{\citenamefont {{Allaria \emph{et al.}}}(2012)}]{allariaetal.HighlyCoherentStable2012}%
  \BibitemOpen
  \bibfield  {author} {\bibinfo {author} {\bibfnamefont {E.}~\bibnamefont {{Allaria \emph{et al.}}}},\ }\bibfield  {title} {\bibinfo {title} {Highly coherent and stable pulses from the {{FERMI}} seeded free-electron laser in the extreme ultraviolet},\ }\href {https://doi.org/10.1038/nphoton.2012.233} {\bibfield  {journal} {\bibinfo  {journal} {Nature Photonics}\ }\textbf {\bibinfo {volume} {6}},\ \bibinfo {pages} {699} (\bibinfo {year} {2012})}\BibitemShut {NoStop}%
\bibitem [{\citenamefont {{Mirian \emph{et al.}}}(2021)}]{mirianetal.GenerationMeasurementIntense2021}%
  \BibitemOpen
  \bibfield  {author} {\bibinfo {author} {\bibfnamefont {N.~S.}\ \bibnamefont {{Mirian \emph{et al.}}}},\ }\bibfield  {title} {\bibinfo {title} {Generation and measurement of intense few-femtosecond superradiant extreme-ultraviolet free-electron laser pulses},\ }\href {https://doi.org/10.1038/s41566-021-00815-w} {\bibfield  {journal} {\bibinfo  {journal} {Nature Photonics}\ }\textbf {\bibinfo {volume} {15}},\ \bibinfo {pages} {523} (\bibinfo {year} {2021})}\BibitemShut {NoStop}%
\bibitem [{\citenamefont {{Nandi \emph{et al.}}}(2022)}]{nandietal.ObservationRabiDynamics2022}%
  \BibitemOpen
  \bibfield  {author} {\bibinfo {author} {\bibfnamefont {S.}~\bibnamefont {{Nandi \emph{et al.}}}},\ }\bibfield  {title} {\bibinfo {title} {Observation of {{Rabi}} dynamics with a short-wavelength free-electron laser},\ }\href {https://doi.org/10.1038/s41586-022-04948-y} {\bibfield  {journal} {\bibinfo  {journal} {Nature}\ }\textbf {\bibinfo {volume} {608}},\ \bibinfo {pages} {488} (\bibinfo {year} {2022})}\BibitemShut {NoStop}%
\bibitem [{\citenamefont {{Richter \emph{et al.}}}(2024)}]{richteretal.StrongfieldQuantumControl2024}%
  \BibitemOpen
  \bibfield  {author} {\bibinfo {author} {\bibfnamefont {F.}~\bibnamefont {{Richter \emph{et al.}}}},\ }\bibfield  {title} {\bibinfo {title} {Strong-field quantum control in the extreme ultraviolet domain using pulse shaping},\ }\href {https://doi.org/10.1038/s41586-024-08209-y} {\bibfield  {journal} {\bibinfo  {journal} {Nature}\ }\textbf {\bibinfo {volume} {636}},\ \bibinfo {pages} {337} (\bibinfo {year} {2024})}\BibitemShut {NoStop}%
\bibitem [{\citenamefont {Olofsson}\ and\ \citenamefont {Dahlstr{\"o}m}(2023)}]{olofssonPhotoelectronSignatureDressedatom2023}%
  \BibitemOpen
  \bibfield  {author} {\bibinfo {author} {\bibfnamefont {E.}~\bibnamefont {Olofsson}}\ and\ \bibinfo {author} {\bibfnamefont {J.~M.}\ \bibnamefont {Dahlstr{\"o}m}},\ }\bibfield  {title} {\bibinfo {title} {Photoelectron signature of dressed-atom stabilization in an intense {{XUV}} field},\ }\href {https://doi.org/10.1103/PhysRevResearch.5.043017} {\bibfield  {journal} {\bibinfo  {journal} {Physical Review Research}\ }\textbf {\bibinfo {volume} {5}},\ \bibinfo {pages} {043017} (\bibinfo {year} {2023})}\BibitemShut {NoStop}%
\bibitem [{Note1()}]{Note1}%
  \BibitemOpen
  \bibinfo {note} {While we will not present results on the following, we should briefly mention that we have found that 2-PRO can indeed be driven in hydrogen between the 1s and 3d. The degeneracy of the 3s and 3d is lifted by the field, allowing one to selectively target the 3d state \cite {bruhnkeTwophotonRabiOscillations2024}}\BibitemShut {NoStop}%
\bibitem [{\citenamefont {Chu}\ and\ \citenamefont {Telnov}(2004)}]{chuFloquetTheoremGeneralized2004}%
  \BibitemOpen
  \bibfield  {author} {\bibinfo {author} {\bibfnamefont {S.-I.}\ \bibnamefont {Chu}}\ and\ \bibinfo {author} {\bibfnamefont {D.~A.}\ \bibnamefont {Telnov}},\ }\bibfield  {title} {\bibinfo {title} {Beyond the {{Floquet}} theorem: Generalized {{Floquet}} formalisms and quasienergy methods for atomic and molecular multiphoton processes in intense laser fields},\ }\href {https://doi.org/10.1016/j.physrep.2003.10.001} {\bibfield  {journal} {\bibinfo  {journal} {Physics Reports}\ }\textbf {\bibinfo {volume} {390}},\ \bibinfo {pages} {1} (\bibinfo {year} {2004})}\BibitemShut {NoStop}%
\bibitem [{\citenamefont {Chu}\ and\ \citenamefont {Reinhardt}(1977)}]{chuIntenseFieldMultiphoton1977}%
  \BibitemOpen
  \bibfield  {author} {\bibinfo {author} {\bibfnamefont {S.-I.}\ \bibnamefont {Chu}}\ and\ \bibinfo {author} {\bibfnamefont {W.~P.}\ \bibnamefont {Reinhardt}},\ }\bibfield  {title} {\bibinfo {title} {Intense {{Field Multiphoton Ionization}} via {{Complex Dressed States}}: {{Application}} to the {{H Atom}}},\ }\href {https://doi.org/10.1103/PhysRevLett.39.1195} {\bibfield  {journal} {\bibinfo  {journal} {Physical Review Letters}\ }\textbf {\bibinfo {volume} {39}},\ \bibinfo {pages} {1195} (\bibinfo {year} {1977})}\BibitemShut {NoStop}%
\bibitem [{\citenamefont {Maquet}\ \emph {et~al.}(1983)\citenamefont {Maquet}, \citenamefont {Chu},\ and\ \citenamefont {Reinhardt}}]{maquetStarkIonizationDc1983}%
  \BibitemOpen
  \bibfield  {author} {\bibinfo {author} {\bibfnamefont {A.}~\bibnamefont {Maquet}}, \bibinfo {author} {\bibfnamefont {S.-I.}\ \bibnamefont {Chu}},\ and\ \bibinfo {author} {\bibfnamefont {W.~P.}\ \bibnamefont {Reinhardt}},\ }\bibfield  {title} {\bibinfo {title} {Stark ionization in dc and ac fields: {{An L}}{\textsuperscript{2}} complex-coordinate approach},\ }\href {https://doi.org/10.1103/PhysRevA.27.2946} {\bibfield  {journal} {\bibinfo  {journal} {Physical Review A}\ }\textbf {\bibinfo {volume} {27}},\ \bibinfo {pages} {2946} (\bibinfo {year} {1983})}\BibitemShut {NoStop}%
\bibitem [{\citenamefont {Suzuki}\ and\ \citenamefont {Okamoto}(1983)}]{suzukiDegeneratePerturbationTheory1983}%
  \BibitemOpen
  \bibfield  {author} {\bibinfo {author} {\bibfnamefont {K.}~\bibnamefont {Suzuki}}\ and\ \bibinfo {author} {\bibfnamefont {R.}~\bibnamefont {Okamoto}},\ }\bibfield  {title} {\bibinfo {title} {Degenerate {{Perturbation Theory}} in {{Quantum Mechanics}}},\ }\href {https://doi.org/10.1143/PTP.70.439} {\bibfield  {journal} {\bibinfo  {journal} {Progress of Theoretical Physics}\ }\textbf {\bibinfo {volume} {70}},\ \bibinfo {pages} {439} (\bibinfo {year} {1983})}\BibitemShut {NoStop}%
\bibitem [{\citenamefont {Foresman}\ \emph {et~al.}(1992)\citenamefont {Foresman}, \citenamefont {{Head-Gordon}}, \citenamefont {Pople},\ and\ \citenamefont {Frisch}}]{foresmanSystematicMolecularOrbital1992}%
  \BibitemOpen
  \bibfield  {author} {\bibinfo {author} {\bibfnamefont {J.~B.}\ \bibnamefont {Foresman}}, \bibinfo {author} {\bibfnamefont {M.}~\bibnamefont {{Head-Gordon}}}, \bibinfo {author} {\bibfnamefont {J.~A.}\ \bibnamefont {Pople}},\ and\ \bibinfo {author} {\bibfnamefont {M.~J.}\ \bibnamefont {Frisch}},\ }\bibfield  {title} {\bibinfo {title} {Toward a systematic molecular orbital theory for excited states},\ }\href {https://doi.org/10.1021/j100180a030} {\bibfield  {journal} {\bibinfo  {journal} {The Journal of Physical Chemistry}\ }\textbf {\bibinfo {volume} {96}},\ \bibinfo {pages} {135} (\bibinfo {year} {1992})}\BibitemShut {NoStop}%
\bibitem [{\citenamefont {Dreuw}\ and\ \citenamefont {{Head-Gordon}}(2005)}]{dreuwSingleReferenceInitioMethods2005}%
  \BibitemOpen
  \bibfield  {author} {\bibinfo {author} {\bibfnamefont {A.}~\bibnamefont {Dreuw}}\ and\ \bibinfo {author} {\bibfnamefont {M.}~\bibnamefont {{Head-Gordon}}},\ }\bibfield  {title} {\bibinfo {title} {Single-{{Reference}} ab {{Initio Methods}} for the {{Calculation}} of {{Excited States}} of {{Large Molecules}}},\ }\href {https://doi.org/10.1021/cr0505627} {\bibfield  {journal} {\bibinfo  {journal} {Chemical Reviews}\ }\textbf {\bibinfo {volume} {105}},\ \bibinfo {pages} {4009} (\bibinfo {year} {2005})}\BibitemShut {NoStop}%
\bibitem [{\citenamefont {Simon}(1979)}]{simonDefinitionMolecularResonance1979}%
  \BibitemOpen
  \bibfield  {author} {\bibinfo {author} {\bibfnamefont {B.}~\bibnamefont {Simon}},\ }\bibfield  {title} {\bibinfo {title} {The definition of molecular resonance curves by the method of exterior complex scaling},\ }\href {https://doi.org/10.1016/0375-9601(79)90165-8} {\bibfield  {journal} {\bibinfo  {journal} {Physics Letters A}\ }\textbf {\bibinfo {volume} {71}},\ \bibinfo {pages} {211} (\bibinfo {year} {1979})}\BibitemShut {NoStop}%
\bibitem [{\citenamefont {Rohringer}\ \emph {et~al.}(2006)\citenamefont {Rohringer}, \citenamefont {Gordon},\ and\ \citenamefont {Santra}}]{rohringerConfigurationinteractionbasedTimedependentOrbital2006}%
  \BibitemOpen
  \bibfield  {author} {\bibinfo {author} {\bibfnamefont {N.}~\bibnamefont {Rohringer}}, \bibinfo {author} {\bibfnamefont {A.}~\bibnamefont {Gordon}},\ and\ \bibinfo {author} {\bibfnamefont {R.}~\bibnamefont {Santra}},\ }\bibfield  {title} {\bibinfo {title} {Configuration-interaction-based time-dependent orbital approach for ab initio treatment of electronic dynamics in a strong optical laser field},\ }\href {https://doi.org/10.1103/PhysRevA.74.043420} {\bibfield  {journal} {\bibinfo  {journal} {Physical Review A}\ }\textbf {\bibinfo {volume} {74}},\ \bibinfo {pages} {043420} (\bibinfo {year} {2006})}\BibitemShut {NoStop}%
\bibitem [{\citenamefont {Greenman}\ \emph {et~al.}(2010)\citenamefont {Greenman}, \citenamefont {Ho}, \citenamefont {Pabst}, \citenamefont {Kamarchik}, \citenamefont {Mazziotti},\ and\ \citenamefont {Santra}}]{greenmanImplementationTimedependentConfigurationinteraction2010}%
  \BibitemOpen
  \bibfield  {author} {\bibinfo {author} {\bibfnamefont {L.}~\bibnamefont {Greenman}}, \bibinfo {author} {\bibfnamefont {P.~J.}\ \bibnamefont {Ho}}, \bibinfo {author} {\bibfnamefont {S.}~\bibnamefont {Pabst}}, \bibinfo {author} {\bibfnamefont {E.}~\bibnamefont {Kamarchik}}, \bibinfo {author} {\bibfnamefont {D.~A.}\ \bibnamefont {Mazziotti}},\ and\ \bibinfo {author} {\bibfnamefont {R.}~\bibnamefont {Santra}},\ }\bibfield  {title} {\bibinfo {title} {Implementation of the time-dependent configuration-interaction singles method for atomic strong-field processes},\ }\href {https://doi.org/10.1103/PhysRevA.82.023406} {\bibfield  {journal} {\bibinfo  {journal} {Physical Review A}\ }\textbf {\bibinfo {volume} {82}},\ \bibinfo {pages} {023406} (\bibinfo {year} {2010})}\BibitemShut {NoStop}%
\bibitem [{\citenamefont {Bertolino}\ \emph {et~al.}(2022)\citenamefont {Bertolino}, \citenamefont {Carlstr{\"o}m}, \citenamefont {Peschel}, \citenamefont {Zapata}, \citenamefont {Lindroth},\ and\ \citenamefont {Dahlstr{\"o}m}}]{bertolinoThomasReicheKuhnCorrectionTruncated2022}%
  \BibitemOpen
  \bibfield  {author} {\bibinfo {author} {\bibfnamefont {M.}~\bibnamefont {Bertolino}}, \bibinfo {author} {\bibfnamefont {S.}~\bibnamefont {Carlstr{\"o}m}}, \bibinfo {author} {\bibfnamefont {J.}~\bibnamefont {Peschel}}, \bibinfo {author} {\bibfnamefont {F.}~\bibnamefont {Zapata}}, \bibinfo {author} {\bibfnamefont {E.}~\bibnamefont {Lindroth}},\ and\ \bibinfo {author} {\bibfnamefont {J.~M.}\ \bibnamefont {Dahlstr{\"o}m}},\ }\bibfield  {title} {\bibinfo {title} {Thomas--{{Reiche--Kuhn}} correction for truncated configuration-interaction spaces: {{Case}} of laser-assisted dynamical interference},\ }\href {https://doi.org/10.1103/PhysRevA.106.043108} {\bibfield  {journal} {\bibinfo  {journal} {Physical Review A}\ }\textbf {\bibinfo {volume} {106}},\ \bibinfo {pages} {043108} (\bibinfo {year} {2022})}\BibitemShut {NoStop}%
\bibitem [{\citenamefont {Kobe}\ and\ \citenamefont {Smirl}(1978)}]{kobeGaugeInvariantFormulation1978}%
  \BibitemOpen
  \bibfield  {author} {\bibinfo {author} {\bibfnamefont {D.~H.}\ \bibnamefont {Kobe}}\ and\ \bibinfo {author} {\bibfnamefont {A.~L.}\ \bibnamefont {Smirl}},\ }\bibfield  {title} {\bibinfo {title} {Gauge invariant formulation of the interaction of electromagnetic radiation and matter},\ }\href {https://doi.org/10.1119/1.11264} {\bibfield  {journal} {\bibinfo  {journal} {American Journal of Physics}\ }\textbf {\bibinfo {volume} {46}},\ \bibinfo {pages} {624} (\bibinfo {year} {1978})}\BibitemShut {NoStop}%
\bibitem [{\citenamefont {Moiseyev}(2011)}]{moiseyevNonHermitianQuantumMechanics2011}%
  \BibitemOpen
  \bibfield  {author} {\bibinfo {author} {\bibfnamefont {N.}~\bibnamefont {Moiseyev}},\ }\href@noop {} {\emph {\bibinfo {title} {Non-{{Hermitian Quantum Mechanics}}}}}\ (\bibinfo  {publisher} {Cambridge University Press},\ \bibinfo {year} {2011})\BibitemShut {NoStop}%
\bibitem [{See()}]{SeeSupplementalMaterial}%
  \BibitemOpen
  \href@noop {} {\bibinfo {title} {See {{Supplemental Material}} at {\emph{[}}{{{\emph{URL}}}}{\emph{ will be inserted by publisher]}} for a brief introduction to {{Floquet}} theory, its application to the semiclassical two-level system, and analytical expressions for the dynamics of the damped effective two-level system. {{See}} also {{Ref}}. [51-52] therein.}}\BibitemShut {Stop}%
\bibitem [{\citenamefont {Gu{\'e}rin}\ \emph {et~al.}(1997)\citenamefont {Gu{\'e}rin}, \citenamefont {Monti}, \citenamefont {Dupont},\ and\ \citenamefont {Jauslin}}]{guerinRelationCavitydressedStates1997}%
  \BibitemOpen
  \bibfield  {author} {\bibinfo {author} {\bibfnamefont {S.}~\bibnamefont {Gu{\'e}rin}}, \bibinfo {author} {\bibfnamefont {F.}~\bibnamefont {Monti}}, \bibinfo {author} {\bibfnamefont {J.-M.}\ \bibnamefont {Dupont}},\ and\ \bibinfo {author} {\bibfnamefont {H.~R.}\ \bibnamefont {Jauslin}},\ }\bibfield  {title} {\bibinfo {title} {On the relation between cavity-dressed states, {{Floquet}} states, {{RWA}} and semiclassical models},\ }\href {https://doi.org/10.1088/0305-4470/30/20/020} {\bibfield  {journal} {\bibinfo  {journal} {Journal of Physics A: Mathematical and General}\ }\textbf {\bibinfo {volume} {30}},\ \bibinfo {pages} {7193} (\bibinfo {year} {1997})}\BibitemShut {NoStop}%
\bibitem [{\citenamefont {Shore}(1990)}]{shoreTheoryCoherentAtomic1990}%
  \BibitemOpen
  \bibfield  {author} {\bibinfo {author} {\bibfnamefont {B.~W.}\ \bibnamefont {Shore}},\ }\href@noop {} {\emph {\bibinfo {title} {The {{Theory}} of {{Coherent Atomic Excitation}}}}}\ (\bibinfo  {publisher} {Wiley VCH},\ \bibinfo {address} {New York},\ \bibinfo {year} {1990})\BibitemShut {NoStop}%
\bibitem [{\citenamefont {Killingbeck}\ and\ \citenamefont {Jolicard}(2003)}]{killingbeckBlochWaveOperator2003}%
  \BibitemOpen
  \bibfield  {author} {\bibinfo {author} {\bibfnamefont {J.~P.}\ \bibnamefont {Killingbeck}}\ and\ \bibinfo {author} {\bibfnamefont {G.}~\bibnamefont {Jolicard}},\ }\bibfield  {title} {\bibinfo {title} {The {{Bloch}} wave operator: Generalizations and applications: {{Part I}}. {{The}} time-independent case},\ }\href {https://doi.org/10.1088/0305-4470/36/20/201} {\bibfield  {journal} {\bibinfo  {journal} {Journal of Physics A: Mathematical and General}\ }\textbf {\bibinfo {volume} {36}},\ \bibinfo {pages} {R105} (\bibinfo {year} {2003})}\BibitemShut {NoStop}%
\bibitem [{\citenamefont {Lindgren}\ and\ \citenamefont {Morrison}(1982)}]{lindgrenAtomicManyBodyTheory1982}%
  \BibitemOpen
  \bibfield  {author} {\bibinfo {author} {\bibfnamefont {I.}~\bibnamefont {Lindgren}}\ and\ \bibinfo {author} {\bibfnamefont {J.}~\bibnamefont {Morrison}},\ }\href {https://doi.org/10.1007/978-3-642-96614-9} {\emph {\bibinfo {title} {Atomic {{Many-Body Theory}}}}},\ edited by\ \bibinfo {editor} {\bibfnamefont {V.~I.}\ \bibnamefont {Goldanskii}}, \bibinfo {editor} {\bibfnamefont {R.}~\bibnamefont {Gomer}}, \bibinfo {editor} {\bibfnamefont {F.~P.}\ \bibnamefont {Sch{\"a}fer}},\ and\ \bibinfo {editor} {\bibfnamefont {J.~P.}\ \bibnamefont {Toennies}},\ \bibinfo {series} {Springer {{Series}} in {{Chemical Physics}}}, Vol.~\bibinfo {volume} {13}\ (\bibinfo  {publisher} {Springer},\ \bibinfo {address} {Berlin, Heidelberg},\ \bibinfo {year} {1982})\BibitemShut {NoStop}%
\bibitem [{\citenamefont {{Allaria \emph{et al.}}}(2013)}]{allariaetal.TwostageSeededSoftXray2013}%
  \BibitemOpen
  \bibfield  {author} {\bibinfo {author} {\bibfnamefont {E.}~\bibnamefont {{Allaria \emph{et al.}}}},\ }\bibfield  {title} {\bibinfo {title} {Two-stage seeded soft-{{X-ray}} free-electron laser},\ }\href {https://doi.org/10.1038/nphoton.2013.277} {\bibfield  {journal} {\bibinfo  {journal} {Nature Photonics}\ }\textbf {\bibinfo {volume} {7}},\ \bibinfo {pages} {913} (\bibinfo {year} {2013})}\BibitemShut {NoStop}%
\bibitem [{\citenamefont {{Finetti \emph{et al.}}}(2017)}]{finettietal.PulseDurationSeeded2017}%
  \BibitemOpen
  \bibfield  {author} {\bibinfo {author} {\bibfnamefont {P.}~\bibnamefont {{Finetti \emph{et al.}}}},\ }\bibfield  {title} {\bibinfo {title} {Pulse {{Duration}} of {{Seeded Free-Electron Lasers}}},\ }\href {https://doi.org/10.1103/PhysRevX.7.021043} {\bibfield  {journal} {\bibinfo  {journal} {Physical Review X}\ }\textbf {\bibinfo {volume} {7}},\ \bibinfo {pages} {021043} (\bibinfo {year} {2017})}\BibitemShut {NoStop}%
\bibitem [{\citenamefont {{Maroju \emph{et al.}}}(2020)}]{marojuetal.AttosecondPulseShaping2020}%
  \BibitemOpen
  \bibfield  {author} {\bibinfo {author} {\bibfnamefont {P.~K.}\ \bibnamefont {{Maroju \emph{et al.}}}},\ }\bibfield  {title} {\bibinfo {title} {Attosecond pulse shaping using a seeded free-electron laser},\ }\href {https://doi.org/10.1038/s41586-020-2005-6} {\bibfield  {journal} {\bibinfo  {journal} {Nature}\ }\textbf {\bibinfo {volume} {578}},\ \bibinfo {pages} {386} (\bibinfo {year} {2020})}\BibitemShut {NoStop}%
\bibitem [{\citenamefont {Rudner}\ and\ \citenamefont {Lindner}(2020)}]{rudnerBandStructureEngineering2020}%
  \BibitemOpen
  \bibfield  {author} {\bibinfo {author} {\bibfnamefont {M.~S.}\ \bibnamefont {Rudner}}\ and\ \bibinfo {author} {\bibfnamefont {N.~H.}\ \bibnamefont {Lindner}},\ }\bibfield  {title} {\bibinfo {title} {Band structure engineering and non-equilibrium dynamics in {{Floquet}} topological insulators},\ }\href {https://doi.org/10.1038/s42254-020-0170-z} {\bibfield  {journal} {\bibinfo  {journal} {Nature Reviews Physics}\ }\textbf {\bibinfo {volume} {2}},\ \bibinfo {pages} {229} (\bibinfo {year} {2020})}\BibitemShut {NoStop}%
\bibitem [{\citenamefont {Oka}\ and\ \citenamefont {Kitamura}(2019)}]{okaFloquetEngineeringQuantum2019}%
  \BibitemOpen
  \bibfield  {author} {\bibinfo {author} {\bibfnamefont {T.}~\bibnamefont {Oka}}\ and\ \bibinfo {author} {\bibfnamefont {S.}~\bibnamefont {Kitamura}},\ }\bibfield  {title} {\bibinfo {title} {Floquet {{Engineering}} of {{Quantum Materials}}},\ }\href {https://doi.org/10.1146/annurev-conmatphys-031218-013423} {\bibfield  {journal} {\bibinfo  {journal} {Annual Review of Condensed Matter Physics}\ }\textbf {\bibinfo {volume} {10}},\ \bibinfo {pages} {387} (\bibinfo {year} {2019})}\BibitemShut {NoStop}%
\bibitem [{\citenamefont {Bruhnke}(2024)}]{bruhnkeTwophotonRabiOscillations2024}%
  \BibitemOpen
  \bibfield  {author} {\bibinfo {author} {\bibfnamefont {J.}~\bibnamefont {Bruhnke}},\ }\emph {\bibinfo {title} {Two-Photon {{Rabi}} Oscillations in Hydrogen: {{A}} Theoretical Study of Effective {{Hamiltonian}} Approaches}},\ \href@noop {} {Master's thesis},\ \bibinfo  {school} {Lund University} (\bibinfo {year} {2024})\BibitemShut {NoStop}%
\bibitem [{\citenamefont {Brion}\ \emph {et~al.}(2007)\citenamefont {Brion}, \citenamefont {Pedersen},\ and\ \citenamefont {M{\o}lmer}}]{brionAdiabaticEliminationLambda2007}%
  \BibitemOpen
  \bibfield  {author} {\bibinfo {author} {\bibfnamefont {E.}~\bibnamefont {Brion}}, \bibinfo {author} {\bibfnamefont {L.~H.}\ \bibnamefont {Pedersen}},\ and\ \bibinfo {author} {\bibfnamefont {K.}~\bibnamefont {M{\o}lmer}},\ }\bibfield  {title} {\bibinfo {title} {Adiabatic elimination in a lambda system},\ }\href {https://doi.org/10.1088/1751-8113/40/5/011} {\bibfield  {journal} {\bibinfo  {journal} {Journal of Physics A: Mathematical and Theoretical}\ }\textbf {\bibinfo {volume} {40}},\ \bibinfo {pages} {1033} (\bibinfo {year} {2007})}\BibitemShut {NoStop}%
\bibitem [{\citenamefont {Faisal}(1987)}]{faisalTheoryMultiphotonProcesses1987}%
  \BibitemOpen
  \bibfield  {author} {\bibinfo {author} {\bibfnamefont {F.~H.~M.}\ \bibnamefont {Faisal}},\ }\href@noop {} {\emph {\bibinfo {title} {Theory of {{Multiphoton Processes}}}}}\ (\bibinfo  {publisher} {Springer},\ \bibinfo {address} {New York},\ \bibinfo {year} {1987})\BibitemShut {NoStop}%
\bibitem [{\citenamefont {Paulisch}\ \emph {et~al.}(2014)\citenamefont {Paulisch}, \citenamefont {Rui}, \citenamefont {Ng},\ and\ \citenamefont {Englert}}]{paulischAdiabaticEliminationHierarchy2014}%
  \BibitemOpen
  \bibfield  {author} {\bibinfo {author} {\bibfnamefont {V.}~\bibnamefont {Paulisch}}, \bibinfo {author} {\bibfnamefont {H.}~\bibnamefont {Rui}}, \bibinfo {author} {\bibfnamefont {H.~K.}\ \bibnamefont {Ng}},\ and\ \bibinfo {author} {\bibfnamefont {B.-G.}\ \bibnamefont {Englert}},\ }\bibfield  {title} {\bibinfo {title} {Beyond adiabatic elimination: {{A}} hierarchy of approximations for multi-photon processes},\ }\href {https://doi.org/10.1140/epjp/i2014-14012-8} {\bibfield  {journal} {\bibinfo  {journal} {The European Physical Journal Plus}\ }\textbf {\bibinfo {volume} {129}},\ \bibinfo {pages} {12} (\bibinfo {year} {2014})}\BibitemShut {NoStop}%
\bibitem [{\citenamefont {Suzuki}\ and\ \citenamefont {Lee}(1980)}]{suzukiConvergentTheoryEffective1980}%
  \BibitemOpen
  \bibfield  {author} {\bibinfo {author} {\bibfnamefont {K.}~\bibnamefont {Suzuki}}\ and\ \bibinfo {author} {\bibfnamefont {S.~Y.}\ \bibnamefont {Lee}},\ }\bibfield  {title} {\bibinfo {title} {Convergent {{Theory}} for {{Effective Interaction}} in {{Nuclei}}},\ }\href {https://doi.org/10.1143/PTP.64.2091} {\bibfield  {journal} {\bibinfo  {journal} {Progress of Theoretical Physics}\ }\textbf {\bibinfo {volume} {64}},\ \bibinfo {pages} {2091} (\bibinfo {year} {1980})}\BibitemShut {NoStop}%
\bibitem [{\citenamefont {Krenciglowa}\ and\ \citenamefont {Kuo}(1974)}]{krenciglowaConvergenceEffectiveHamiltonian1974}%
  \BibitemOpen
  \bibfield  {author} {\bibinfo {author} {\bibfnamefont {E.~M.}\ \bibnamefont {Krenciglowa}}\ and\ \bibinfo {author} {\bibfnamefont {T.~T.~S.}\ \bibnamefont {Kuo}},\ }\bibfield  {title} {\bibinfo {title} {Convergence of effective hamiltonian expansion and partial summations of folded diagrams},\ }\href {https://doi.org/10.1016/0375-9474(74)90184-5} {\bibfield  {journal} {\bibinfo  {journal} {Nuclear Physics A}\ }\textbf {\bibinfo {volume} {235}},\ \bibinfo {pages} {171} (\bibinfo {year} {1974})}\BibitemShut {NoStop}%
\bibitem [{\citenamefont {{Virtanen \emph{et al.}}}(2020)}]{virtanenetal.SciPy10Fundamental2020}%
  \BibitemOpen
  \bibfield  {author} {\bibinfo {author} {\bibfnamefont {P.}~\bibnamefont {{Virtanen \emph{et al.}}}},\ }\bibfield  {title} {\bibinfo {title} {{{SciPy}} 1.0: Fundamental algorithms for scientific computing in {{Python}}},\ }\href {https://doi.org/10.1038/s41592-019-0686-2} {\bibfield  {journal} {\bibinfo  {journal} {Nature Methods}\ }\textbf {\bibinfo {volume} {17}},\ \bibinfo {pages} {261} (\bibinfo {year} {2020})}\BibitemShut {NoStop}%
\bibitem [{\citenamefont {Weiser}\ and\ \citenamefont {Zarantonello}(1988)}]{weiserNotePiecewiseLinear1988}%
  \BibitemOpen
  \bibfield  {author} {\bibinfo {author} {\bibfnamefont {A.}~\bibnamefont {Weiser}}\ and\ \bibinfo {author} {\bibfnamefont {S.~E.}\ \bibnamefont {Zarantonello}},\ }\bibfield  {title} {\bibinfo {title} {A note on piecewise linear and multilinear table interpolation in many dimensions},\ }\href {https://doi.org/10.1090/S0025-5718-1988-0917826-0} {\bibfield  {journal} {\bibinfo  {journal} {Mathematics of Computation}\ }\textbf {\bibinfo {volume} {50}},\ \bibinfo {pages} {189} (\bibinfo {year} {1988})}\BibitemShut {NoStop}%
\bibitem [{\citenamefont {Suzuki}(1982)}]{suzukiConstructionHermitianEffective1982}%
  \BibitemOpen
  \bibfield  {author} {\bibinfo {author} {\bibfnamefont {K.}~\bibnamefont {Suzuki}},\ }\bibfield  {title} {\bibinfo {title} {Construction of {{Hermitian Effective Interaction}} in {{Nuclei}}: --- {{General Relation}} between {{Hermitian}} and {{Non-Hermitian Forms}} ---},\ }\href {https://doi.org/10.1143/PTP.68.246} {\bibfield  {journal} {\bibinfo  {journal} {Progress of Theoretical Physics}\ }\textbf {\bibinfo {volume} {68}},\ \bibinfo {pages} {246} (\bibinfo {year} {1982})}\BibitemShut {NoStop}%
\end{thebibliography}%

\onecolumngrid

\section*{End Matter}

\twocolumngrid

\section{{Non-Hermitian quantum theory}}

Quantum theory deals with Hermitian operators that have real eigenvalues and orthonormal eigenfunctions. If complex scaling is introduced, the Hamiltonian $H$ of the system becomes non-Hermitian, with complex eigenvalues $E_n - \i \Gamma_n$ and complex-scaled eigenfunctions $\ket{n}$ that are not orthonormal under the usual scalar product \cite{moiseyevNonHermitianQuantumMechanics2011}. However, the eigenfunctions still fulfill bi-orthonormality $\braket{\tilde m| n} = \delta_{nm}$, where $\bra{\tilde m}$ solves the left eigenvalue problem $\bra{\tilde m} H = (E_n - \i \Gamma_n) \bra{\tilde m}$. The closure relation then reads $1 = \sum_n \ket{n}\bra{\tilde n}$. Note that with this notation, we do not differentiate if the non-Hermitian $H$ is symmetric ($H_F^\theta$) or non-symmetric ($H_\eff$).

\section{Effective Hamiltonian theory}

\subsection{{Energy-dependent effective Hamiltonian}}

The purpose of an effective Hamiltonian $H_\eff$ is to reproduce a subset of the true eigenvalues and eigenstate projections of the full Hamiltonian. We can obtain such an object easily by partitioning the time-independent Schrödinger equation (TISE) $H\ket{\Psi} = E\ket{\Psi}$ \cite{lindgrenAtomicManyBodyTheory1982}. Let the time-independent full Hamiltonian $H$ of our system be non-Hermitian. Suppose we can make a perturbative ansatz, $H = H_0 + V$, where $H_0 \ket{n} = E_n \ket{n}$. Let us now select a small subset of $N$ unperturbed essential states $\ket{n}$ that span an $N$-dimensional model space $\mathcal{P}$. We project onto $\mathcal{P}$ via the projection operator $P = \sum_{n \in \P} \ketbra{n}{\tilde n}$. The $\mathcal{Q}$-space is the space spanned by the non-essential states, i.e. $\Q = \P^\perp$, where $Q = 1 - P$ projects onto $\mathcal{Q}$. From this construction, it follows that $[H_0, P] = [H_0, Q] = 0$, as well as $PQ = QP = 0$, $P^2 = P$, and $Q^2 = Q$. 

Starting with the TISE, we insert $P+Q = 1$ before $\ket{\Psi}$, and then project from the left with either $P$ or $Q$:
\begin{align}
    PHP \ket{\Psi} + PHQ \ket{\Psi} &= E P \ket{\Psi} \label{eq:PprojTISE}\\
    QHQ \ket{\Psi} + QHP \ket{\Psi} &= E Q \ket{\Psi} \label{eq:QprojTISE}
\end{align}
Since $[H_0, P] = [H_0, Q] = 0$, we have $PH_0Q = QH_0P = 0$, leading to $PHQ = PVQ$ and $QHP = QVP$. Rearranging Eq.~\eqref{eq:QprojTISE} to $Q\ket{\Psi}$ and inserting into Eq.~\eqref{eq:PprojTISE}, we obtain a TISE for the $\P$-space dynamics, 
\begin{align}
    \underbrace{\left( PHP + PVQ \frac{Q}{E - QHQ} QVP \right)}_{\eqqcolon H_\eff(E)} P\ket{\Psi} = E P \ket{\Psi}, \label{eq:energydep_TISE}
\end{align}
with the caveat that the effective Hamiltonian $H_\eff(E)$ in $\P$ is energy-dependent. Hence, Eq.~\eqref{eq:energydep_TISE} constitutes a non-linear eigenvalue problem and must be solved self-consistently one eigenenergy at a time \cite{lindgrenAtomicManyBodyTheory1982}.

\subsection{{Energy-independent effective Hamiltonian}}

An energy-independent effective Hamiltonian reproduces \textit{all} desired eigenvalues and eigenstate projections simultaneously. Obtaining such an object is the subject of effective Hamiltonian theory \cite{killingbeckBlochWaveOperator2003}. In this work, we use a first-order approximation for the energy-independent effective Hamiltonian. This keeps the formalism conceptually simple, yet as our results demonstrate, sufficiently sophisticated.\par 

The energy-dependence of $H_\eff(E)$ is weak if the $\P$-space eigenvalues are close to another and well-separated from the $\Q$-space spectrum. In this case, a suitable approximation is to evaluate $H_\eff(E)$ at a fixed energy $E_0$ that lies approximately in the center of the $\P$-space spectrum. This is the pole approximation \cite{cohen-tannoudjiAtomPhotonInteractionsBasic1998}, yielding the zeroth-order (energy-independent) effective Hamiltonian
\begin{equation}
    H_\eff^{(0)} = H_\eff(E_0) = PHP + PVQ \frac{Q}{E_0 - QHQ} QVP. \label{eq:Heff0}
\end{equation}
$H_\eff^{(0)}$ is equivalent to the effective Hamiltonian from adiabatic elimination \cite{brionAdiabaticEliminationLambda2007}. In studies of resonant ionisation using effective Hamiltonians, $H_\eff^{(0)}$ is the standard choice \cite{beersExactSolutionRealistic1975, olofssonPhotoelectronSignatureDressedatom2023}. Often, $[E_0 - QHQ]^{-1} \equiv [E_0 - QH_0Q - QVQ]^{-1}$ is expanded perturbatively for small $V$ \cite{cohen-tannoudjiAtomPhotonInteractionsBasic1998}, enabling convenient implementation of the slowly-varying envelope approximation for smooth pulse envelopes \cite{faisalTheoryMultiphotonProcesses1987}. 

For strong perturbations, the eigenstates of the full Hamiltonian will be perturbed significantly, so that the gap between the $\P$-space and $\Q$-space spectrum decreases. Then, the energy-dependence of $H_\eff(E)$ can become non-negligible near the $\P$-space eigenvalues and the pole approximation becomes invalid. A first-order correction to the pole approximation, Eq.~\eqref{eq:Heff0}, can be obtained by linearly expanding $[E - QHQ]^{-1}$ around $E_0$:
\begin{align}
    H_\eff(E) &\approx H_\eff^{(0)} - (E - E_0) \cdot \mathcal{C}, \label{eq:HeffE_linear}
\end{align}
where
\begin{align}
\mathcal{C} &\coloneqq PVQ \frac{Q}{(E_0 - QHQ)^2} QVP. \label{eq:correction}
\end{align}
Inserting into the $\P$-space TISE, Eq.~\eqref{eq:energydep_TISE}, and rearranging yields the energy-independent
\begin{align}
    H_\eff^{(1)} &= (P + \mathcal{C})^{-1} (H_\eff^{(0)} + E_0\mathcal{C} ).\label{eq:Heff1}
\end{align}
When $E_0 = 0$, $H_\eff^{(1)}$ is equivalent to the first-order effective Hamiltonian obtained by Paulisch \textit{et al.} using higher-order Markov approximations \cite{paulischAdiabaticEliminationHierarchy2014}, revealing the analogy between the energy and time domains in effective Hamiltonian theory. Another more formal angle is to interpret $H_\eff^{(0)}$ as the first term in the Lee-Suzuki iterative sequence, and $H_\eff^{(1)}$ as the second term \cite{suzukiConvergentTheoryEffective1980}. The eigenvalues of $H_\eff^{(1)}$ approximately reproduce a subset of the true eigenvalues of the full system, $H \ket{\Psi_n} = \lambda_n \ket{\Psi_n}$, and the eigenstates of $H_\eff^{(1)}$ approximately reproduce the corresponding eigenstate projections $P\ket{\Psi_n}$. We note here that a range of iterative and perturbative procedures exist for the calculation of converged effective Hamiltonians, see e.g. Ref.~\cite{krenciglowaConvergenceEffectiveHamiltonian1974, suzukiConvergentTheoryEffective1980, lindgrenAtomicManyBodyTheory1982}. 

In this work, we use Eq.~\eqref{eq:Heff1} to calculate the effective Hamiltonian, which we simply abbreviate as $H_\eff$. We obtain the inverses $[E-QHQ]^{-k}$, $k=1,2$, numerically exact via LU decomposition. This is computationally cheap because the Floquet basis in our calculations typically contains between 1000 and 5000 atom-photon states. For Fig.~\ref{fig:poprabi} and~\ref{fig:resonancecondition}, we interpolate between effective Hamiltonians $H_\eff^{(1)}$ in the parameter space spanned by $\omega$ and $I_0$. The cubic interpolation is performed using the \texttt{RegularGridInterpolator} class from SciPy \cite{virtanenetal.SciPy10Fundamental2020, weiserNotePiecewiseLinear1988}. For hydrogen (1s-2s transition), we sample from five frequencies and intensities each. In helium, it depends on the transition. For 1s$^2$-1s2p, we sample from nine frequencies and ten intensities. For 1s$^2$-1s2s, we sample from eight frequencies and ten intensities. For 1s$^2$-1s3d and 1s$^2$-1s4d, we sample from ten frequencies and intensities each. This yields $H_\eff^{(1)}(\omega, I_0)$. Fig.~\ref{fig:poprabi}, \ref{fig:resonancecondition}, and~\ref{fig:CRoscillations} are then calculated through upcoming analytical formulas. 

If $H$ is complex-symmetric, then $H_\eff^{(0)}$ will also be complex-symmetric, while $H_\eff^{(1)}$ will generally be non-symmetric. Analogously, if $H$ is Hermitian, then $H_\eff^{(0)}$ will be Hermitian and $H_\eff^{(1)}$ will generally be non-Hermitian. This property of higher-order effective Hamiltonians is well-known \cite{suzukiDegeneratePerturbationTheory1983}: The projected eigenstates $P\ket{\Psi_n}$ are generally not orthogonal to each other, $\Braket{ \Psi_n | P |\Psi_m} \neq \delta_{nm}$, despite the full eigenstates being orthogonal: $\Braket{\Psi_n|\Psi_m} = \delta_{nm}$ (in the complex-scaled case, this argument applies for bi-orthogonality). 

We note here that there also exists a complex-symmetric effective Hamiltonian $H_\effcs^{(1)}$ with the same eigenvalues as $H_\eff^{(1)}$, but orthonormal eigenvectors in the c-product. $H_\effcs^{(1)}$ can be obtained from $H_\eff^{(1)}$ via similarity transform with $(P+\mathcal{C})^{1/2}$. While $H_\eff^{(1)}$ and $H_\effcs^{(1)}$ give rise to the same physical observables \cite{suzukiConstructionHermitianEffective1982}, we are forced to work with $H_\eff^{(1)}$ when comparing populations between TDCIS and effective Hamiltonian, since we need to work in the same basis in both approaches.\par\bigskip

\subsection{{Reduced wave operator}}

Applying the pole approximation,
\begin{align}
    \chi \coloneqq \frac{Q}{E_0 - QHQ} QVP, \label{eq:reducedwaveop}
\end{align}
gives an estimate for the $\Q$-space wavefunction based on the $\P$-space wavefunction,
\begin{equation}
    Q\ket{\Psi} \simeq \chi P \ket{\Psi}. \label{eq:QfromPwavefunction}
\end{equation}
It thus approximates the reduced wave operator from time-independent perturbation theory (also called the correlation operator) \cite{lindgrenAtomicManyBodyTheory1982}. Note how $\chi$ from Eq.~\eqref{eq:reducedwaveop} emerges as a by-product when calculating $H_\eff^{(0)}$ or $H_\eff^{(1)}$. It involves no further computational effort. 

We use $\chi$ in this work in order to model the $\Q$-space time-evolution that is due to the essential dressed energies. This enables us to rigorously model counter-rotating effects, since these require knowledge of the $\mathcal{Q}$-space components of the dressed states. In particular, for an essential state $\ket{b,-2}$ and the CR states $\ket{b,0}$ and $\ket{b,-4}$, the characteristic quantity for the prediction of CR oscillations is given by
\begin{equation}
    \Lambda \coloneqq \braket{b,0|\chi|b,-2} + \braket{b,-4|\chi|b,-2}.
\end{equation}

\end{document}